\documentclass[twocolumn,aps,floats,letterpaper,floatfix,groupedaddress]{revtex4-1}
\usepackage{amsmath}
\usepackage{graphicx}
\usepackage{amsfonts}
\usepackage{amssymb}
\usepackage{bm}
\usepackage{epsfig,float,afterpage}
\usepackage[colorlinks,linkcolor=blue,citecolor=blue,urlcolor=blue]{hyperref}
\usepackage[usenames,dvipsnames]{color}

\newcommand{\beq}{\begin{equation}}
\newcommand{\eeq}{\end{equation}}
\newcommand{\bes}{\begin{subequations}}
\newcommand{\ees}{\end{subequations}}
\newcommand{\bea}{\begin{eqnarray}}
\newcommand{\eea}{\end{eqnarray}}
\newcommand{\ba}{\begin{array}}
\newcommand{\ea}{\end{array}}
\newcommand{\beqn}{\begin{eqnarray*}}
\newcommand{\eeqn}{\end{eqnarray*}}

\newcommand{\f}[2]{\frac{#1}{#2}}

\newcommand{\om}{\omega}

\newcommand{\la}{\langle}
\newcommand{\ra}{\rangle}

\newcommand{\dg}{\dagger}
\newcommand{\Om}{\Omega}
\newcommand\scalemath[2]{\scalebox{#1}{\mbox{\ensuremath{\displaystyle #2}}}}

\def\nn{\nonumber}

\begin{document}
\title{Amplification and cross-Kerr nonlinearity in waveguide quantum electrodynamics}
\author{Athul Vinu and Dibyendu Roy}
\affiliation{Raman Research Institute, Bangalore 560080, India}
\begin{abstract}
  We explore amplification and cross-Kerr nonlinearity by a three-level emitter (3LE) embedded in a waveguide and driven by two light beams. The coherent amplification and cross-Kerr nonlinearity were demonstrated in recent experiments, respectively, with a $V$ and a ladder-type 3LE coupled to an open superconducting transmission line carrying two microwave fields.  Here, we consider $\Lambda,V$, and ladder-type 3LE, and compare the efficiency of coherent and incoherent amplification as well as the magnitude of the cross-Kerr phase shift in all three emitters. We apply the Heisenberg-Langevin equations approach to investigate the scattering of a probe and a drive beam both initially in a coherent state. We particularly calculate the regime of the probe and drive powers when the 3LE acts most efficiently as a coherent amplifier, and derive the second-order coherence of amplified probe photons. Finally, we apply the Kramers-Kronig relations to correlate the amplitude and phase response of the probe beam, which are used in finding the coherent amplification and the cross-Kerr phase shift in these systems.   
\end{abstract}

\maketitle
\section{Introduction}
Waveguide quantum electrodynamics (QED) systems \cite{RoyRMP2017, Gu2017} are a new platform for investigating the coherent and incoherent scattering of few propagating photons from individual atoms embedded in a one-dimensional (1D) waveguide. Superconducting quantum circuits \cite{RoyRMP2017,Gu2017}, tapered nanofibers \cite{Petersen14}, and photonic crystals \cite{LodahlRMP2015} are some examples of such systems. Strong light-matter interactions have been engineered in these waveguide QED systems to demonstrate many interesting physical phenomena such as resonance fluorescence \cite{ShenFanPRL2007, Astafiev10a, Zheng10}, nonreciprocal transmission \cite{RoyPRB2010, RoyNatS2013, Mitsch14, FratiniPRL2014, RoyPRA2017, Hamann2018}, electromagnetically induced transparency \cite{Abdumalikov10,Witthaut10,RoyPRL2011,Roy2014}, cross-Kerr nonlinearity \cite{He2011, HoiPRL2013}, photon-mediated interactions between distant emitters \cite{Zheng13a,vanLoo2013}, quantum wave mixing \cite{Dmitriev2017,Honigl2018}, and to create basic all-optical quantum devices such as single-photon router or switch \cite{Abdumalikov10,Hoi11,Shomroni2014}, single-photon transistor \cite{Hwang09,Bajcsy2009}, amplifier \cite{AstafievPRL2010, Oelsner2013, Koshino2013, Shevchenko2014,Wen2018}.

Many of the above phenomena, e.g., electromagnetically induced transparency, cross-Kerr nonlinearity, nonreciprocity,  and the devices, e.g., router, transistor, amplifier, are studied with a three-level emitter (3LE) and two light beams. Depending on the used optical transitions for the two beams, various configurations of the 3LE are employed. For example, \textcite{AstafievPRL2010} implemented on-chip quantum amplification of a probe beam on a single V-type 3LE by creating population inversion using a drive beam. \textcite{HoiPRL2013} realized a cross-Kerr interaction between two microwave fields by strongly coupling a ladder-type 3LE to an open superconducting transmission line carrying the microwave fields. A comparison of optimal gain for coherent amplification in different configurations of 3LE for a weak probe and a strong drive beam was carried out in Ref.~\cite{ZhaoPRA2017}. While many studies \cite{AstafievPRL2010,ZhaoPRA2017} discuss on-chip coherent amplification of a probe beam, the coherent amplification in these systems is accompanied by an incoherent amplification, which has been mostly ignored so far. The incoherent amplification limits the performance of coherent amplification and severely controls the statistics of the amplified probe beam. The coherent amplification is needed for linear (phase-sensitive) amplifiers and the total amplification including coherent and incoherent amplification can be useful for photon detectors including single-photon detectors. 

In the first part of this paper, we perform a detailed analysis of coherent and  incoherent amplification in different models of 3LE for arbitrary strength of probe and pump beams. In superconducting circuits, a flux qubit \cite{AstafievPRL2010}, a transmon qubit \cite{HoiPRL2013}, and a heavy fluxonium qubit (a capacitively shunted fluxonium circuit) \cite{Earnest2018} can be used to realize respectively a $V$, a ladder and a $\Lambda$-type 3LE. We apply here the Heisenberg-Langevin equations approach \cite{Koshino2012,RoyPRA2017,Manasi2018} to investigate the time-evolution of light fields and the emitter after their interaction. We begin the results by arguing that only $\Lambda$ and V configurations of the 3LE can amplify, and then discuss why a $\Lambda$-type 3LE acts a better amplifier than a $V$-type 3LE at low drive power.  We derive approximate formulas for coherent and incoherent  amplification, which show the dependence of these on drive power and inelastic (non-radiative) losses. We compare between coherent and incoherent amplification at different probe and drive power, and point out where the on-chip device acts most efficiently as a coherent amplifier. We particularly  show that the maximum coherent amplification is much higher in a V system than a $\Lambda$ system for a resonant weak probe beam at substantial drive powers. We also calculate second-order coherence $g^{(2)}(\tau)$ (with delay time $\tau$) of amplified probe photons. For a relatively  strong drive field, we find $g^{(2)}(\tau=0)<1$ at low probe powers when the incoherent amplification dominates and $g^{(2)}(\tau=0) \ge 1$ at  higher probe powers when the coherent amplification is significant.     

While the amplitude response of a transmitted probe field in the presence of a drive field gives a measure of the coherent amplification, a difference in the  phase response of the transmitted probe field in the presence and absence of a drive field is used to quantify cross-Kerr interaction between the probe and drive fields. We examine cross-Kerr phase shifts in all three different 3LEs. We find that though the above definition to quantify the cross-Kerr phase shift works perfectly well for a $V$ and a ladder system, we need to introduce a different description to quantify the cross-Kerr phase shift in a $\Lambda$ system accurately. We calculate the cross-Kerr phase shift as a function of the power of probe and drive beam at a probe frequency that maximizes the phase shift. Such dependences of the cross-Kerr phase shift on power were measured in a ladder system by \textcite{HoiPRL2013}, and our results are in agreement with the experiment. Finally, we apply the Kramers-Kronig relations in understanding the connection between the coherent amplification and the cross-Kerr phase shift of the probe beam in these systems. We primarily identify a regime of probe powers when a measurement of amplitude response of probe transmission as a function of probe beam detuning can be used to derive the exact phase response of the probe beam.

The rest of the paper is arranged in the following sections. In Sec.~\ref{model}, we introduce the Hamiltonians of different 3LEs, photon and excitation fields, and interactions between them. We describe the Heisenberg-Langevin equations approach to calculate transport properties in $\Lambda$-type 3LE in Sec.~\ref{trans}. We discuss the coherent and incoherent amplification of a probe beam in a $\Lambda$ and a $V$ system in Sec.~\ref{amp}. The investigation of the cross-Kerr phase shift in all three emitters and the relation between amplitude and phase response of a probe beam are given in Sec.~\ref{cross}. We conclude our article with a discussion in Sec.~\ref{dis}. We have further added six Appendixes at the end to include the linearization of photon dispersion in our Hamiltonian, the details of the Heisenberg-Langevin equations approach to derive transport properties in an $\Lambda$-type system, the transport properties in $V$- and ladder-type 3LEs, and a comparison between quantum and classical modeling of the drive beam. 

\section{Models and Hamiltonians}
\label{model}
\begin{figure}
\includegraphics[width=\columnwidth]{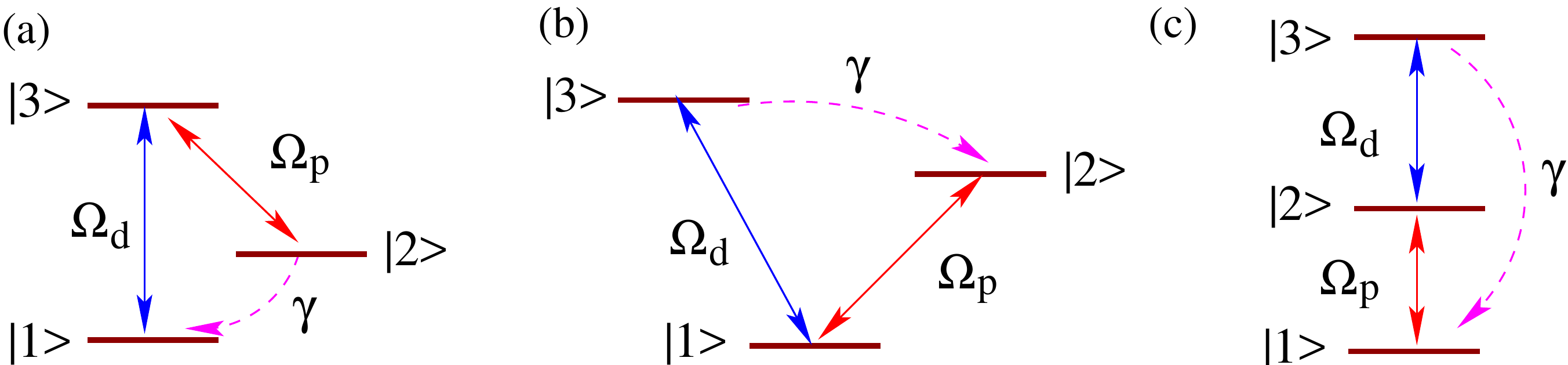}
\caption{Cartoon of 3LEs with levels $|1\ra,|2\ra,|3\ra$. Three different arrangements for the coupling of probe (red arrow) and drive (blue arrow) beams with the allowed transitions make (a) $\Lambda$, (b) $V$, and (c) ladder-type configuration of the 3LE. Here, $\Omega_p$ and $\Omega_d$ are respectively the Rabi frequency of the probe and drive beam, and $\gamma$ denotes the strength of non-radiative decay.}
\label{3le}
\end{figure}

We consider a 3LE with energy levels $|1\ra,|2\ra$, and $|3\ra$. We set the energy of the ground level $|1\ra$ to be zero. The energies of the levels $|2\ra$ and $|3\ra$ are respectively $\hbar\om_{21}$ and $\hbar \om_{31}$ with $\om_{21}<\om_{31}$. The 3LE is embedded in an open 1D waveguide. There are total three different optical transitions between levels $|1\ra \leftrightarrow |2\ra$, $|2\ra \leftrightarrow |3\ra$, and $|3\ra \leftrightarrow |1\ra$. A monochromatic, continuous-wave beam of frequency $\om_d$ drives one of the transitions of the 3LE, and we choose $\om_d$ close to that particular transition frequency. Another monochromatic, continuous-wave probe beam of frequency $\om_p$ is side-coupled to another transition of the 3LE. Depending on the used optical transitions for the drive and probe beams, we classify the 3LE as $V$, $\Lambda$, and ladder-type system (see Fig.~\ref{3le}). The third allowed transition of these systems can be connected by an additional coupling field or a non-radiative relaxation process forming $\Delta$-type cyclic transitions. Here, we consider the presence of a relaxation channel, e.g., a non-radiative decay  at the third transition for creating a population inversion \cite{AstafievPRL2010,HoiPRL2013}. We also assume here that the polarization of the drive and probe beam is different. It helps us to separate scattered probe and drive lights.

For a full quantum modeling, we express both the probe and drive beams as quantum electromagnetic fields, and write the Hamiltonian of an X-type of 3LE in a 1D waveguide as:
\bea
\f{\mathcal{H}_X^q}{\hbar}&=&\om_{21} \sigma^{\dg}\sigma+ \om_{31}\mu^{\dg}\mu+\int_{-\infty}^{\infty}dk \Big(v_gk \big[\sum_{\alpha=\pm}(a_{k\alpha}^{\dg}a_{k\alpha}\nn\\&-&b_{k\alpha}^{\dg}b_{k\alpha})+c_{k}^{\dg}c_{k}+d_{k}^{\dg}d_{k}+f_{k}^{\dg}f_{k}\big]+\lambda_2(c_k+c_k^{\dg})\sigma^{\dg}\sigma\nn\\&+&\lambda_3(f_k+f_k^{\dg})\mu^{\dg}\mu\Big)+ \f{\mathcal{H}^c_X}{\hbar},\label{Ham1Q}
\eea
where $X$ stands for $\Lambda$ or $V$ or ladder-type 3LE. The raising and lowering operators of the emitter are defined as, $\sigma^{\dg}\equiv |2\ra \la 1|, \sigma \equiv |1\ra \la 2|, \mu^{\dg}\equiv |3\ra \la 2|, \mu \equiv |2\ra \la 3|, \nu^{\dg} \equiv |1\ra \la 3|$ and $\nu \equiv |3\ra \la 1|$. $a_{k\alpha}^{\dg}~[b_{k\alpha}^{\dg}]$ are creation operators for two different polarizations of right-moving [left-moving] photon modes of the probe and drive beams. Here, the polarizations are denoted by subscript $\alpha=\pm$, and we assign $+$ and $-$ polarization respectively for the probe and drive beam. $d_k^{\dg}$ is a creation operator of the excitations related to non-radiative decay. The strength of pure dephasing of levels $|2\ra$ and $|3\ra$ to baths of excitations created by operators $c_k^{\dg}$ and $f_k^{\dg}$ are respectively $\lambda_2$ and $\lambda_3$. We assume a linear energy-momentum dispersion, $\om_k=v_g k$,  near some frequencies which are close to the relevant optical transitions, and $v_g$ is the group velocity of photons at those frequencies (check Appendix~\ref{AppLin}). $\mathcal{H}^c_X$ consists of coupling of the probe and drive beams with the 3LE, and it also contains the interaction of 3LE with the relaxation process.

We here write light-matter interactions in linear form (dipole approximation) within the rotating-wave approximation. The coupling strength of the probe and drive beam  with the 3LE is respectively $g_p$ and $g_d$. Within the Markov approximation, the couplings $g_p$ and $g_d$ are taken to be constant over photon frequency near the corresponding optical transitions. For such frequency independent couplings, the photon fields behave as memoryless baths. Therefore, we write for a $\Lambda,V$ and ladder ($\Xi$)-type 3LE:
\bea
\f{\mathcal{H}^c_{\Lambda}}{\hbar}&=&\int_{-\infty}^{\infty}dk \big[g_p\mu^{\dg}\beta_{k+}+g_d\nu \beta_{k-}+\gamma \sigma^{\dg} d_k+h.c.\big], \label{HamLam}\\
\f{\mathcal{H}^c_{V}}{\hbar}&=&\int_{-\infty}^{\infty}dk \big[g_p\sigma^{\dg}\beta_{k+}+g_d\nu \beta_{k-}+\gamma \mu^{\dg} d_k+h.c.\big],\label{HamV} \\
\f{\mathcal{H}^c_{\Xi}}{\hbar}&=&\int_{-\infty}^{\infty}dk \big[g_p\sigma^{\dg}\beta_{k+}+g_d\mu^{\dg}\beta_{k-}+\gamma \nu d_k+h.c.\big],\label{HamLad}
\eea
where we define $\beta_{k\pm}=(a_{k\pm}+b_{k\pm})$, and $\gamma$ denotes a strength of coupling to the non-radiative decay. 

\section{Transport properties}\label{3}
\label{trans}
Here, we consider both the input beams from the left of the waveguide QED system. The input probe and drive beams are in the coherent states with frequency $\om_p$ and $\om_d$, and with amplitude $E_p$ and $E_d$, respectively. We assume that the bath of non-radiative decay is initially in vacuum mode at $t=t_0$. Therefore, the full initial state is $|\psi\ra=|E_p,\om_p\ra \otimes |E_d,\om_d\ra \otimes |\varphi\ra$ which satisfies $a_{k+}(t_0)|\psi \ra=E_p\delta (v_gk-\om_p)| \psi \ra$, $a_{k-}(t_0)|\psi \ra=E_d\delta (v_gk-\om_d)| \psi \ra$ and $b_{k+}(t_0)| \psi \ra=b_{k-}(t_0)| \psi \ra=0$. Here, $|\varphi\ra$ is the vacuum of the bath of non-radiative decay and pure-dephasing, and we have $c_k(t_0)|\varphi\ra=d_k(t_0)|\varphi\ra=f_k(t_0)|\varphi\ra=0$. We assume that the couplings $g_p$ and $g_d$ are turned on at $t=t_0$ when the light beams are shined on the 3LE in the ground state $|1\ra$. Following \textcite{AstafievPRL2010}, we set pure dephasing rates $\Gamma_{\lambda2}=\pi\lambda_2^2/v_g$ and $\Gamma_{\lambda3}=\pi\lambda_3^2/v_g$ to zero in our following discussion of amplification \footnote{The presence of pure-dephasing affects coherent amplification more significantly than incoherent amplification.}.

Below, we provide details of calculation for the $\Lambda$-type 3LE, and the corresponding results for the $V$ and ladder systems are given in Appendix~\ref{App1} and \ref{App2}, respectively. We apply the Heisenberg-Langevin equations approach with the Hamiltonian in Eqs.~\ref{Ham1Q} and \ref{HamLam} to calculate the time-evolution of the emitter and light fields after they interact (check Appendix~\ref{AppHL} for a derivation). Thus, we derive the following matrix equations for the time-evolution of emitter's operators after taking their expectation in the initial state $|\psi\ra$ of the light fields:
\begin{widetext}
\bea
\f{d\boldsymbol{\mathcal{M}}_{\Lambda}}{dt}=\boldsymbol{\mathcal{R}}_{\Lambda}\boldsymbol{\mathcal{M}}_{\Lambda}+\boldsymbol{\Omega}_{\Lambda},~~{\rm where}~~\boldsymbol{\mathcal{R}}_{\Lambda}=
&&\left( \begin{array}{cccccccc} \kappa_1  & -i\Omega_p & 0 & i\Omega_d & -i\Omega_d & 0 & 0 & 0 \\ -i\Omega_p & \kappa_3^* & i\Omega_d & 0 & 0 & 0 & 0 & 0 \\ 0 & i\Omega_d & \kappa_2^*  & -2i\Omega_p & -i\Omega_p & 0 & 0 & 0 \\i\Omega_d   & 0 & -i\Omega_p & -2\tilde{\Gamma} & 0 & i\Omega_p & 0 & -i\Omega_d \\ -i\Omega_d & 0 & 0 & 4\Gamma_d-2\Gamma_{\gamma} & -2\Gamma_{\gamma} & 0 & 0 & i\Omega_d \\ 0 & 0 & 0 & 2i\Omega_p & i\Omega_p & \kappa_2  & -i\Omega_d & 0 \\ 0 & 0 & 0 & 0 & 0 & -i\Omega_d &\kappa_3  & i\Omega_p \\ 0 & 0 & 0 & -i\Omega_d & i\Omega_d & 0 & i\Omega_p & \kappa_1^* \end{array}\right),\label{eom3LEq}
\eea
\end{widetext}
$\boldsymbol{\mathcal{M}}_{\Lambda}(t)=(\mathcal{N}^*_1,\mathcal{S}_1,\mathcal{M}^*_1,\mathcal{N}_2,\mathcal{S}_2,\mathcal{M}_1,\mathcal{S}^*_1,\mathcal{N}_1)^{T}$ and $\boldsymbol{\Omega}_{\Lambda}=(0,0,i\Omega_p,0,2\Gamma_{\gamma},-i\Omega_p, 0, 0)^{T}$. We define the Rabi frequency of the probe beam by $\Omega_p=g_pE_p/v_g$, and identify that of the control beam as $\Omega_d=g_dE_d/v_g$. The detunings of the beams from the respective transitions are $\Delta_d=\omega_{31}-\omega_d$ and $\Delta_p=\omega_{31}-\omega_{21}-\omega_p$. Within the Heisenberg-Langevin equations approach, we derive time-evolution of emitter's operators in Eq.~\ref{eom3LEq} by integrating out the light fields. Such integration out of some parts of the full Hamiltonian $\mathcal{H}_{\Lambda}^q$ induces dissipation and decoherence in the emitter's part of the Hamiltonian. The different relaxation rates in our Eq.~\ref{eom3LEq} are due to such dissipation and decoherence arising from the integration out of different photonic and excitation fields. The relaxation rates arose due to the couplings of each mode (left or right) of probe and drive beams are written respectively as $\Gamma_p=\pi g_p^2/v_g$ and $\Gamma_d=\pi g_d^2/v_g$. The other parameters are non-radiative decay rate $\Gamma_{\gamma}=\pi \gamma^2/v_g$, total relaxation rate due to both left and right-moving drive and probe photons $\tilde{\Gamma}=2(\Gamma_p+\Gamma_d)$ and diagonal entries $\kappa_1=-i\Delta_d-\tilde{\Gamma},~\kappa_2=-i\Delta_p-\tilde{\Gamma}-\Gamma_{\gamma},~\kappa_3=\kappa_2-\kappa_1$. We have used the following definitions for the expectation of the emitter's operators in $\boldsymbol{\mathcal{M}}_{\Lambda}(t)$:
\bea
\mathcal{N}_1(t)&=&\la \psi |\nu(t)| \psi \ra e^{-i\om_d(t-t_0)}, \nn\\
\mathcal{S}_1(t)&=&\la \psi |\sigma(t)| \psi \ra e^{i(\om_d-\om_p)(t-t_0)},\nn\\
\mathcal{S}_2(t)&=&\la \psi|\sigma(t)\sigma^{\dg}(t)| \psi \ra,\nn\\
\mathcal{M}_1(t)&=&\la \psi |\mu(t)| \psi \ra e^{i\om_p(t-t_0)}, \nn\\
\mathcal{N}_2(t)&=&\la \psi |\nu(t)\nu^{\dg}(t)| \psi \ra.\nn
\eea

Using the solution of $\boldsymbol{\mathcal{M}}_{\Lambda}(t)$ from Eq.~\ref{eom3LEq}, we can evaluate the transport coefficients of the probe and drive beams and their power spectra. We introduce a real-space description of the propagating photons at position $x \in [-\infty,\infty]$ to derive the properties of the scattered light \cite{RoyPRA2017}. For the left-moving and right-moving probe and drive photons, we define $a_{x\alpha}(t)=\int_{-\infty}^{\infty} dk\:e^{ikx}a_{k\alpha}(t)/\sqrt{2\pi}$ and $b_{x\alpha}(t)=\int_{-\infty}^{\infty} dk\:e^{ikx}b_{k\alpha}(t)/\sqrt{2\pi}$. Here, the photon operators at $x<0$ and $x>0$ denote respectively the incident and scattered photons, and the photons at $x=0$ are coupled to the emitter.

The power spectrum of light represents a distribution of photons over frequency. For example, the power spectrum of the incident, monochromatic probe light is a delta function around $\om_p$. We define power spectrum of transmitted probe and drive light at long-time steady-state as
\bea
P_{\rm tr,\alpha}(t,\om)={\rm Re}\int_0^{\infty}\f{d\tau}{\pi}e^{i\om \tau}\la a_{x\alpha}^{\dg}(t)a_{x\alpha}(t+\tau)\ra,\label{power1} 
\eea
where we take $x >0, t\gg t_0$ and the expectation $\la..\ra$ is performed in the initial state $|\psi\ra$. We can derive the power spectrum of the incident probe and drive beams, $P_{\rm in, +}(\om)$ and $P_{\rm in, -}(\om)$ respectively, by writing expressions like Eq.~\ref{power1} for $a_{x\alpha}(t)$ at $x <0$. As expected, we find $P_{\rm in, +}(\om)=E_p^2\delta(\om-\om_p)/(2\pi v_g^2)$ and $P_{\rm in, -}(\om)=E_d^2\delta(\om-\om_d)/(2\pi v_g^2)$. We get the total incident probe and drive power from the integrated power spectra: $\int d\om P_{\rm in,+}(\om)=E_p^2/(2\pi v_g^2)\equiv I_{\rm p}$ and $\int d\om P_{\rm in,-}(\om)=E_d^2/(2\pi v_g^2)\equiv I_{\rm d}$. The power spectrum of the reflected probe and drive lights at long-time steady-state is 
\bea
P_{\rm ref,\alpha}(t,\om)={\rm Re}\int_0^{\infty}\f{d\tau}{\pi}e^{i\om \tau}\la b_{x\alpha}^{\dg}(t)b_{x\alpha}(t+\tau)\ra,\label{power2} 
\eea
where again we take $x >0, t\gg t_0$.

The transmission coefficients $\mathcal{T}^{\Lambda}_p(t),\mathcal{T}^{\Lambda}_d(t)$ and the reflection coefficients $\mathcal{R}^{\Lambda}_p(t),\mathcal{R}^{\Lambda}_d(t)$ of the probe and drive beams at some time $t$ are derived by dividing the respective integrated power spectrum by the incident power $I_{\rm p}$ or $I_{\rm d}$ (see Appendix~\ref{AppDt} for details):
\bea
\mathcal{T}^{\Lambda}_p(t)&=&1+\f{4\Gamma_p^2}{\Omega_p^2} \mathcal{N}_{2}(t)+\f{4\Gamma_p}{\Omega_p} {\rm Im}[\mathcal{M}_{1}(t)],\label{transp}\\
\mathcal{T}^{\Lambda}_d(t)&=&1+\f{4\Gamma_d^2}{\Omega_d^2} \mathcal{N}_{2}(t)-\f{4\Gamma_d}{\Omega_d} {\rm Im}[\mathcal{N}_{1}(t)],\label{transd}\\
\mathcal{R}^{\Lambda}_p(t)&=&\f{4\Gamma_p^2}{\Omega_p^2} \mathcal{N}_{2}(t),~\mathcal{R}^{\Lambda}_d(t)=\f{4\Gamma_d^2}{\Omega_d^2} \mathcal{N}_{2}(t).\label{refd}
\eea
We can find $\mathcal{N}_{1}(t),\mathcal{N}_{2}(t)$, and $\mathcal{M}_{1}(t)$ by solving Eq.~\ref{eom3LEq}, and calculate the above transport coefficients of light from Eqs.~\ref{transp}-\ref{refd}. The long-time steady-state properties of the light-matter interaction can be obtained from Eq.~\ref{eom3LEq} by setting $\f{d\boldsymbol{\mathcal{M}}_{\Lambda}}{dt}=0$, and we find $\boldsymbol{\mathcal{M}}_{\Lambda}(t \to \infty)=-\boldsymbol{\mathcal{R}}_{\Lambda}^{-1}\boldsymbol{\Omega}_{\Lambda}$. In the following, we apply these transport coefficients in investigating amplification of the probe beam in the presence of a drive beam and non-radiative decay. In Appendix~\ref{App3}, we discuss some differences in the transport properties at long-time steady-state due to quantum and classical modeling of the drive beam when $\Omega_d \to 0$.

\section{Amplification by population inversion}
\label{amp}
We notice that the excitation number operators $N_1$ and $N_2$ commute with $\mathcal{H}_{\Lambda}^{q}$ in the absence of non-radiative decay ($\Gamma_{\gamma}=0$), where
\bea
N_1&=&\sigma^{\dg}\sigma+\mu^{\dg}\mu+\int_{-\infty}^{\infty}dk (a_{k-}^{\dg}a_{k-}+b_{k-}^{\dg}b_{k-}), \\
N_2&=&-\sigma^{\dg}\sigma+\int_{-\infty}^{\infty}dk (a_{k+}^{\dg}a_{k+}+b_{k+}^{\dg}b_{k+}).
\eea
Therefore, these number operators remain conserved at all times during the light-matter interactions. We can further deduce from Eq.~\ref{eom3LEq} that $\boldsymbol{\mathcal{M}}_{\Lambda}(t)$ reaches a unique steady-state value due to the relaxation terms in $\boldsymbol{\mathcal{R}}_{\Lambda}$ via the Markovian light-matter coupling. Thus, $\la \psi|\sigma^{\dg}(t)\sigma(t)| \psi \ra$ and $\la \psi |\mu^{\dg}(t)\mu(t)| \psi \ra$ are independently time-invariant in the steady-state. From the above arguments, we  conclude that $\la \psi|\int_{-\infty}^{\infty}dk (a_{k+}^{\dg}a_{k+}+b_{k+}^{\dg}b_{k+})| \psi \ra$ and $\la \psi|\int_{-\infty}^{\infty}dk (a_{k-}^{\dg}a_{k-}+b_{k-}^{\dg}b_{k-})| \psi \ra$ do not change over time in the steady-state. Therefore, there is no exchange of photons between the different polarization of light fields in the steady-state of a driven $\Lambda$-type 3LE for any value of $\Omega_d$ at $\Gamma_{\gamma}=0$.

For a finite $\Gamma_{\gamma}$, the number operators $N_1$ and $N_2$ do not commute with $\mathcal{H}_{\Lambda}^{q}$. Thus, these are no longer conserved quantities. Nevertheless, we find
\bea
&&[N_1+\int_{-\infty}^{\infty} dk\:d_k^{\dg}d_k,\mathcal{H}_{\Lambda}^{q}]=0,\\
&&[N_2-\int_{-\infty}^{\infty} dk\:d_k^{\dg}d_k,\mathcal{H}_{\Lambda}^{q}]=0.
\eea
Again, we can argue like before that $\la \psi|\sigma^{\dg}(t)\sigma(t)|\psi \ra$ and $\la \psi |\mu^{\dg}(t)\mu(t)| \psi \ra$ are independently constant over time in the steady-state of the driven $\Lambda$-type 3LE. In the steady-state, we can also prove $\la \psi|\int_{-\infty}^{\infty}dk \:d_{k}^{\dg}d_{k}|\psi \ra$ increases linearly with time as the rate of decay of $d\la \psi|\int_{-\infty}^{\infty}dk \:d_{k}^{\dg}d_{k}|\psi \ra/dt$ is $2\Gamma_{\gamma}\la \psi|\sigma^{\dg}(t)\sigma(t) |\psi\ra$ ($2\Gamma_{\gamma}\la \psi|\mu^{\dg}(t)\mu(t) |\psi\ra$) for a $\Lambda$ ($V$) system which is a constant. Therefore, $\la \psi|\int_{-\infty}^{\infty}dk (a_{k-}^{\dg}a_{k-}+b_{k-}^{\dg}b_{k-})| \psi \ra $ decays with time linearly while $\la \psi|\int_{-\infty}^{\infty}dk (a_{k+}^{\dg}a_{k+}+b_{k+}^{\dg}b_{k+})| \psi \ra$ grows linearly with time in the steady-state. The above argument suggests the amplification of probe beam ($+$ polarization) by an exchange of photons between the probe and drive beams via the non-radiative decay in a $\Lambda$-type 3LE. 


We now explicitly quantify the amount of amplification in different types of 3LEs. We first observe that the sum of transmission and reflection of the probe and drive beams are independently conserved in the steady-state when $\Gamma_{\gamma}=0$; thus we have $\mathcal{T}^{\Lambda}_p(t \to \infty)+\mathcal{R}^{\Lambda}_p(t \to \infty)=1$ and $\mathcal{T}^{\Lambda}_d(t \to \infty)+\mathcal{R}^{\Lambda}_d(t \to \infty)=1$. This proves no exchange of photons between the probe and drive beams at steady-state as we have argued earlier. However, there is an exchange of photons between the beams if we include a finite non-radiative decay ($\Gamma_{\gamma}\ne 0$) in our $\Lambda$-type 3LE, and we have  $\mathcal{T}^{\Lambda}_p(t \to \infty)+\mathcal{R}^{\Lambda}_p(t \to \infty)>1,~\mathcal{T}^{\Lambda}_d(t \to \infty)+\mathcal{R}^{\Lambda}_d(t \to \infty)<1$ with the constrain $I_{\rm p}(\mathcal{T}^{\Lambda}_p(t \to \infty)+\mathcal{R}^{\Lambda}_p(t \to \infty))+I_{\rm d}(\mathcal{T}^{\Lambda}_d(t \to \infty)+\mathcal{R}^{\Lambda}_d(t \to \infty))=(I_{\rm p}+I_{\rm d})$. Therefore, the driven $\Lambda$-type 3LE acts as an amplifier for $\Gamma_{\gamma}\ne 0$. The amplification of the probe beam in this system is due to the optical pumping of the emitter's population from the ground to excited state by the drive beam and the cyclic transitions created by the non-radiative decay to continue the process. Thus, the mechanism of coherent amplification here is the well-known population inversion for standard lasing operation. A similar line of argument can be made for a V-type 3LE for the mixing of photons from the probe and control beams via population inversion in the presence of non-radiative decay \footnote{For $V$ system, we have $[N_1+\int_{-\infty}^{\infty} dk\:d_k^{\dg}d_k,\mathcal{H}_{V}^{q}]=[N_2-\int_{-\infty}^{\infty} dk\:d_k^{\dg}d_k,\mathcal{H}_{V}^{q}]=0$ where $N_1=\mu^{\dg}\mu+\int_{-\infty}^{\infty}dk (a_{k-}^{\dg}a_{k-}+b_{k-}^{\dg}b_{k-})$, $N_2=\sigma^{\dg}\sigma+\int_{-\infty}^{\infty}dk (a_{k+}^{\dg}a_{k+}+b_{k+}^{\dg}b_{k+})$}. However, the population inversion can not be created for a ladder-type 3LE, and the previous arguments with the conserved excitation numbers infer a loss of photons from both the probe and drive beams in  the presence of  non-radiative decay \footnote{For $\Xi$ system, we have $[N_1+\int_{-\infty}^{\infty} dk\:d_k^{\dg}d_k,\mathcal{H}_{\Xi}^{q}]=[N_2+\int_{-\infty}^{\infty} dk\:d_k^{\dg}d_k,\mathcal{H}_{\Xi}^{q}]=0$ where $N_1=\mu^{\dg}\mu+\int_{-\infty}^{\infty}dk (a_{k-}^{\dg}a_{k-}+b_{k-}^{\dg}b_{k-})$, $N_2=\sigma^{\dg}\sigma+\mu^{\dg}\mu+\int_{-\infty}^{\infty}dk (a_{k+}^{\dg}a_{k+}+b_{k+}^{\dg}b_{k+})$}. Hereafter, we discuss various features of the amplification in a $\Lambda$ and a $V$-type 3LE.

It is apt here to clarify the role of light-matter interaction and waveguide in amplification and nonlinearity in our study. While the physical mechanism for amplification of a probe beam in our $\Lambda$ and $V$-type 3LE is the population inversion created by a drive beam, a strong light-emitter interaction inside waveguide generates a large amplification by a single on-chip emitter. This is the unique feature of waveguide systems; thus, such on-chip devices are termed as ultimate quantum amplifiers \cite{AstafievPRL2010}. As photon-photon scattering in a vacuum is extremely weak, light-matter interaction is an essential ingredient for interactions between photons (e.g., between the probe and drive photons) or optical nonlinearity. However, light-matter coupling at individual emitter in three-dimensional free-space is relatively small due to spatial-mode mismatch between the incident and scattered electromagnetic waves. The matching problem of spatial modes of light can be overcome by confining light inside waveguide; this increases the efficiency of light-matter coupling substantially. Thus, both the light-matter coupling and the confinement inside waveguide are essential to creating strong optical nonlinearity. This is the physical mechanism for strong cross-Kerr nonlinearity between the probe and drive beams studied in a later section. The cross-Kerr nonlinearity is also part of the amplification process, which is an exchange of photons between the probe and drive beams.

\subsection{Coherent amplification}

\begin{figure}
\includegraphics[width=\columnwidth]{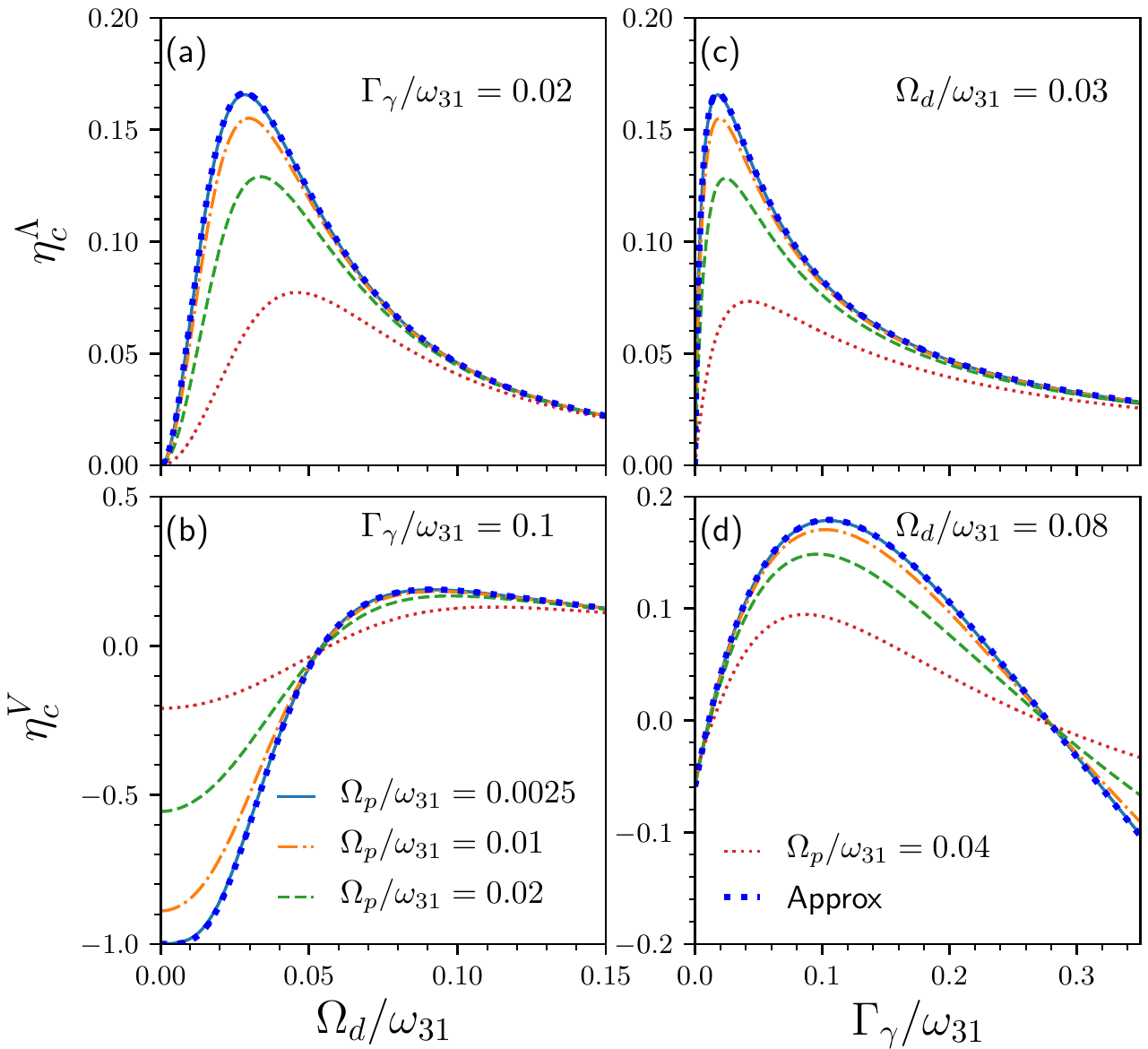}
\caption{Features of coherent amplification $\eta^{\Lambda}_c$ and $\eta^{V}_c$ of a probe beam by respectively a $\Lambda$ and a $V$-type 3LE coupled to a 1D waveguide. (a,b) Scaling of $\eta^{\Lambda}_c$ and $\eta^{V}_c$ as a function of the Rabi frequency $\Omega_d$ of the drive beam, and (c,d) dependence of $\eta^{\Lambda}_c$ and $\eta^{V}_c$ on non-radiative decay $\Gamma_{\gamma}$. The parameters are $\Gamma_p=\Gamma_d=0.01,\Delta_p=\Delta_d=0$ in all panels, and (a) $\Gamma_{\gamma}=0.02$, (c) $\Gamma_{\gamma}=0.1$, (b) $\Om_d=0.03$, (d) $\Om_d=0.08$. The parameters are in unit of $\om_{31}$.}
\label{cohA}
\end{figure}

We first study the amplification of the coherently scattered part of a probe beam. This has been investigated experimentally in waveguide QED using a V-type 3LE made of a superconducting quantum circuit coupled to a 1D transmission line \cite{AstafievPRL2010}. Coherent amplification of a probe beam is also discussed in recent theoretical studies in waveguide QED systems \cite{ZhaoPRA2017}. We define coherent amplification efficiency $\eta_c$ as the ratio of the difference between the coherently transmitted power and incident power, and the incident power itself. $\eta_c$ is a measure of what fraction of the amplified, transmitted probe photons has a constant phase relation with the incident probe beam. The coherent amplification efficiency for a $\Lambda$ and a $V$-type 3LE is respectively given by: 
\bea
\eta^{\Lambda}_c=\f{2\big(\Gamma_p|\mathcal{M}_1(t)|^2+\Omega_p{\rm Im}[\mathcal{M}_1(t)]\big)}{v_gI_{\rm p}},\label{CampL}\\
\eta^{V}_c=\f{2\big(\Gamma_p|\mathcal{S}_3(t)|^2+\Omega_p{\rm Im}[\mathcal{S}_3(t)]\big)}{v_gI_{\rm p}}.\label{CampV}
\eea

When both the probe and drive beams are on resonant, e.g., $\Delta_p=\Delta_d=0$, we can find approximate expressions for the coherent amplification of a weak probe beam. These for the $\Lambda$ and $V$-type 3LE are $\eta^{X}_c \approx 4\eta^{X}_0(\eta^{X}_0+1)$ where $X=\Lambda,V$, and  
\bea
\eta^{\Lambda}_0&=&\f{\Gamma_p\Gamma_{\gamma}\Omega_d^2(2\Gamma_d+\Gamma_{\gamma})}{(\Omega_d^2+\Gamma_t\Gamma_{\gamma})(\tilde{\Gamma}^2\Gamma_{\gamma}+2\Omega_d^2(\Gamma_{\gamma}+\Gamma_p))}, \label{CampLa}\\
\eta^{V}_0&=&\f{\Gamma_p(\Gamma_t(\Gamma_{\gamma}\Omega_d^2-2\Gamma_p(2\Gamma_d+\Gamma_{\gamma})^2)-4\Gamma_p^2\Omega_d^2)}{(2\Gamma_p\Gamma_t+\Omega_d^2)(2\Gamma_p(2\Gamma_d+\Gamma_{\gamma})^2+\Omega_d^2(\Gamma_{\gamma}+4\Gamma_p))},\nn\\\label{CampVa}
\eea
with $\Gamma_t=\tilde{\Gamma}+\Gamma_{\gamma}$. In Fig.~\ref{cohA}(a,b), we plot these approximate formulas of $\eta^{\Lambda}_c$ and $\eta^{V}_c$ from Eqs.~\ref{CampLa},\ref{CampVa} as a function of $\Omega_d$ of the drive beam, and compare these approximate lineshapes with the exact ones from Eqs.~\ref{CampL},\ref{CampV} for different probe amplitude $\Omega_p$ and a fixed $\Gamma_{\gamma}$. As expected, the approximate formulas match with the exact lineshapes only for small $\Omega_p$. It is clear from Eqs.~\ref{CampLa},\ref{CampVa} that $\eta^{\Lambda}_c>0$ for any small but non-zero $\Omega_{d}$ while it requires a large $\Omega_{d}~(>\Omega_{d0}$ where $\Omega_{d0}=\sqrt{2\Gamma_p\Gamma_t}(2\Gamma_d+\Gamma_{\gamma})/(\Gamma_{\gamma}\Gamma_t-4\Gamma_p^2)^{1/2}$ from Eq.~\ref{CampVa}) to have $\eta^{V}_c>0$. Therefore, a $\Lambda$-type 3LE acts as a better ultimate on-chip quantum amplifier than a $V$-type 3LE at a weak driving field when both the probe and  drive fields are at the few-photon quantum regime. A substantial population inversion to level $|3\ra$ with respect to level $|2\ra$ of a $\Lambda$-type 3LE is achieved with much less pumping by the drive beam in comparison to that to level $|2\ra$ with respect to level $|1\ra$ of a $V$-type 3LE. As the population of level $|2\ra$ of a $\Lambda$-type 3LE is essentially near zero due to non-radiative decay to level $|1\ra$, it requires less pumping for creating a population inversion to level $|3\ra$ in a $\Lambda$-type 3LE. On the contrary, at least half of the population of the emitter must be excited from level $|1\ra$ to levels $|2\ra$ and $|3\ra$ of a $V$-type 3LE by the drive beam to generate a population inversion. The population of level $|2\ra$ becomes higher than that of level $|1\ra$ of a $V$ system when $\Omega_d>\Omega_{d0}$.

We also notice from Fig.~\ref{cohA}(a,b) that both $\eta^{\Lambda}_c$ and $\eta^{V}_c$ depend non-monotonically on $\Omega_{d}$. From Eq.~\ref{CampLa} we find that while $\eta^{\Lambda}_c$ grows quadratically of $\Omega_{d}$ at small $\Omega_{d}$, it decays as $\Omega_{d}^{-2}$ at large $\Omega_{d}$. There is no population inversion in the absence of a drive beam ($\Omega_d=0$), and the population inversion grows with increasing $\Omega_{d}$ from zero. However, the maximum population inversion is reached at a finite $\Omega_{d}$, and a further increment in $\Omega_{d}$ causes splitting of levels $|1\ra$ and $|3\ra$, which  leads to detuning of the probe beam. Therefore, the coherent scattering of the probe beam and related amplification efficiency fall with increasing $\Omega_{d}$ beyond some $\Omega_{d}$.

\begin{figure}[h]
\includegraphics[width=\columnwidth]{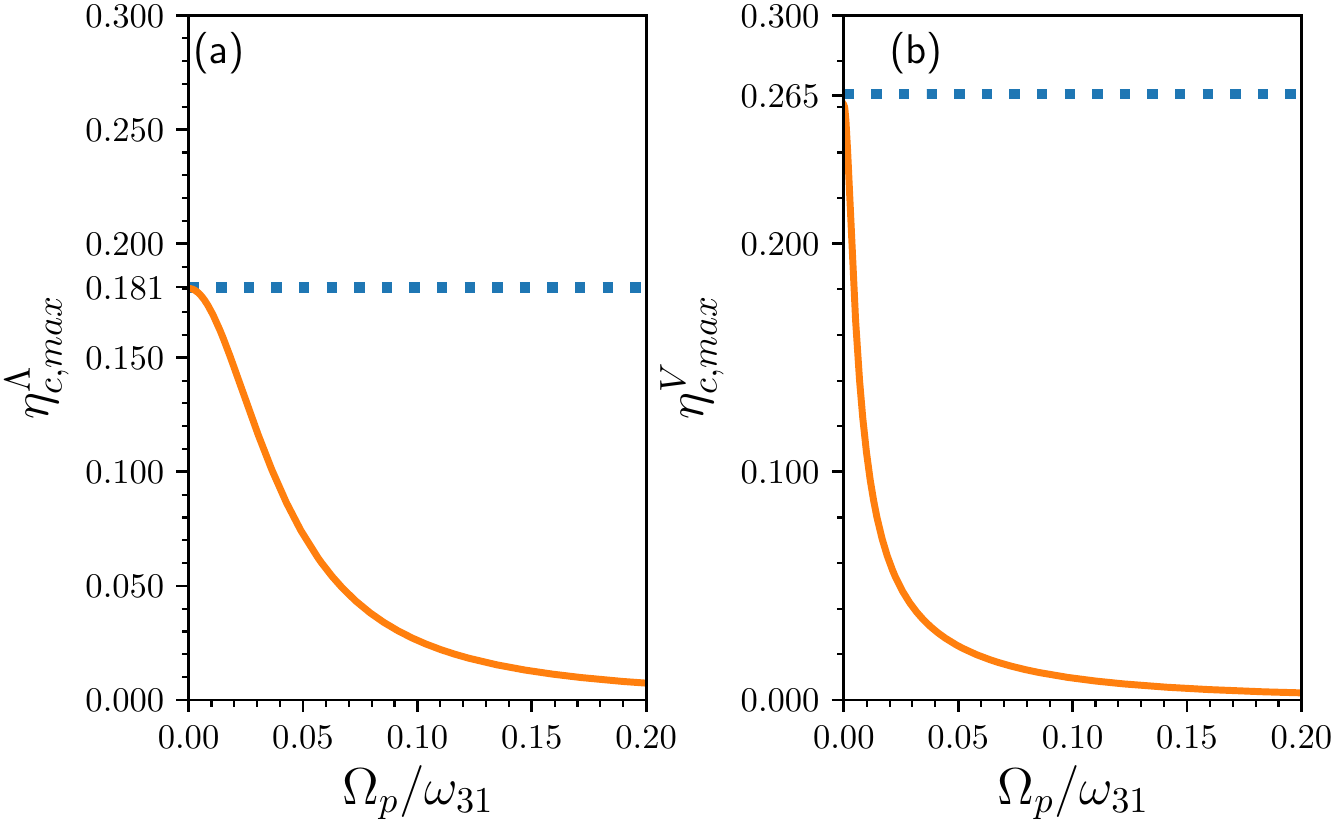}
\caption{Maximum coherent amplification $\eta^{\Lambda}_{c,max}$ and $\eta^{V}_{c,max}$ of a resonant probe beam by respectively a $\Lambda$ and a $V$-type 3LE coupled to a 1D waveguide. Dashed lines are calculated from the approximate formulas in Eqs.~\ref{CampLa},\ref{CampVa} for a weak probe beam, and full lines show exact scaling of $\eta^{\Lambda}_{c,max}$ and $\eta^{V}_{c,max}$ as a function of the Rabi frequency $\Omega_p$ of the probe beam at an $\Omega_d$ that maximizes the coherent amplification. The parameters are $\Gamma_d=0.0001, \Delta_p=\Delta_d=0$ in both panels, and (a) $\Gamma_p=0.01, \Gamma_{\gamma}=0.03$, (b) $\Gamma_p=0.001, \Gamma_{\gamma}=0.1$. The parameters are in unit of $\om_{31}$.}
\label{cohAmax}
\end{figure}

We evaluate the maximum attainable coherent amplification of a weak probe beam in $\Lambda$ and $V$-type 3LEs using the approximate formulas in Eqs.~\ref{CampLa},\ref{CampVa} when both the probe and drive beams are on resonant. For a $\Lambda$-type 3LE, the maximum of $\eta^{\Lambda}_0$ and related $\eta^{\Lambda}_c$ occurs at a critical Rabi frequency of the drive field, $\Omega^2_d|_c=\sqrt{\Gamma_t}\tilde{\Gamma}\Gamma_{\gamma}/\sqrt{2(\Gamma_{\gamma}+\Gamma_p)}$. When we further maximize $\eta^{\Lambda}_0$ at $\Omega^2_d|_c$ and $\Gamma_d \approx 0$ (weak drive-field coupling or low relaxation due to the drive beam), we find the value of maximum possible coherent amplification in a $\Lambda$-type 3LE is 0.181 for $\Gamma_p/\Gamma_{\gamma} \approx 0.351$. To calculate the maximum amount of $\eta^{V}_c$, we approximate $\eta^{V}_0$ in Eq.~\ref{CampVa} further in the regime $\Gamma_d \ll \Gamma_p \ll \Gamma_{\gamma}$ with the limit $\Gamma_d \to 0,\Gamma_t \to \Gamma_{\gamma}$, and we find
\bea
\eta^{V}_0 &\approx& \f{\Gamma_p(\Omega_d^2(\Gamma_{\gamma}^2-4\Gamma_p^2)-2\Gamma_p\Gamma_{\gamma}^3)}{(2\Gamma_p\Gamma_{\gamma}+\Omega_d^2)(2\Gamma_p\Gamma_{\gamma}^2+\Omega_d^2(\Gamma_{\gamma}+4\Gamma_p))} \nn\\
&\approx& \f{1}{2}\f{\nu -1}{(\nu+1)^2},~~~\nu=\f{\Omega_d^2}{2\Gamma_p\Gamma_{\gamma}},
\eea
where we obtain the last line from the previous line by applying $\Gamma_p \ll \Gamma_{\gamma}$ for terms multiplying $\Omega_d^2$. We have a maximum of $\eta^{V}_0$ at $\nu=3$, and the corresponding maximum value of $\eta^{V}_c$ is 0.266. Therefore, we conclude that the coherent amplification of a weak probe beam is higher in a $V$-type 3LE than that in a $\Lambda$-type 3LE when both the drive and probe beams are on resonant. \textcite{AstafievPRL2010} found the maximum value of coherent transmission amplitude $\tilde{t}_p^V$ being $9/8$, which gives the maximum amount of $\eta^{V}_c$ as $|\tilde{t}_p^V|^2-1=17/64$ as before. In Fig.~\ref{cohAmax}, we present the scaling of the maximum value of $\eta^{\Lambda}_c$ and $\eta^{V}_c$ with increasing $\Omega_p$, and also compare them with the maximum value of $\eta^{\Lambda}_c$ and $\eta^{V}_c$ obtained from the approximate formulas in Eqs.~\ref{CampLa},\ref{CampVa} for a weak probe beam. While the maximum value of $\eta^{\Lambda}_c$ and $\eta^{V}_c$ matches with those from the approximate formulas at $\Omega_p \to 0$, they fall with increasing $\Omega_p$ due to saturation of the transition coupled to the probe beam. These are shown in Fig.~\ref{cohAmax}. 

We later explore the role of non-radiative decay rate $\Gamma_{\gamma}$ in coherent amplification. A non-zero $\Gamma_{\gamma}$ is essential for the exchange of photons between the probe and drive beams in the steady-state. Nevertheless, $\Gamma_{\gamma}$ also reduces the coherence of level $|2\ra$ of a $\Lambda$-type 3LE, which would affect coherent amplification. The population of levels $|3\ra$, as well as $|2\ra$ of a $V$-type 3LE, decreases with increasing $\Gamma_{\gamma}$ at larger $\Gamma_{\gamma}$; it leads to a reduction in population inversion in a $V$-type 3LE. Thus, we expect coherent amplification to initially improve with increasing $\Gamma_{\gamma}$ and then to fall beyond certain values of $\Gamma_{\gamma}$ in both $\Lambda$ and $V$-type 3LE. We show the dependence of $\eta^{\Lambda}_c$ and $\eta^{V}_c$ on $\Gamma_{\gamma}$ in Fig.~\ref{cohA}(c,d) which behavior matches with our above arguments

\subsection{Incoherent amplification}

\begin{figure}
\includegraphics[width=\columnwidth]{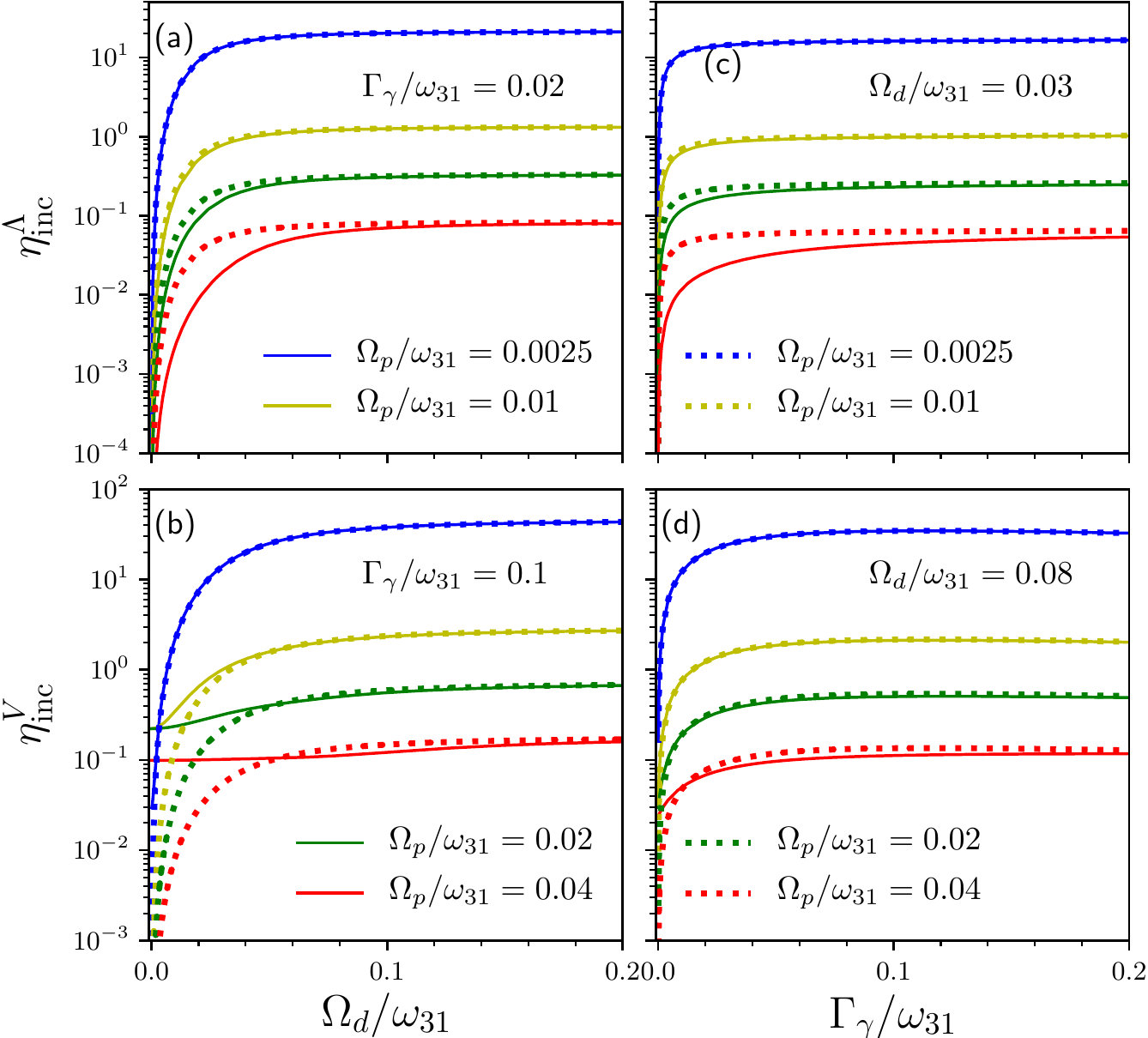}
\caption{Features of incoherent amplification $\eta^{\Lambda}_{\rm inc}$ and $\eta^{V}_{\rm inc}$ of a probe beam by respectively a $\Lambda$ and a $V$-type 3LE coupled to a 1D waveguide. (a,b) Scaling of $\eta^{\Lambda}_{\rm inc}$ and $\eta^{V}_{\rm inc}$ as a function of the Rabi frequency $\Omega_d$ of the drive beam, and (c,d) dependence of $\eta^{\Lambda}_{\rm inc}$ and $\eta^{V}_{\rm inc}$ on non-radiative decay $\Gamma_{\gamma}$. The dotted curves are obtained from the approximate formulas in Eqs.~\ref{IampLa},\ref{IampVa}, and the full lines are using the exact formulas in Eqs.~\ref{IampL},\ref{IampV}. The parameters are $\Gamma_p=\Gamma_d=0.01$ in all panels, and (a) $\Gamma_{\gamma}=0.02$, (c) $\Gamma_{\gamma}=0.1$, (b) $\Om_d=0.03$, (d) $\Om_d=0.08$. The parameters are in unit of $\om_{31}$.}
\label{incohA}
\end{figure}

Next, we consider the amplification of the incoherently scattered part of the probe beam. For this, we first quantify the total amplification efficiency of the coherently and incoherently scattered parts of the probe beam as 
\bea
\eta^{X}_T=\mathcal{T}^{X}_p(t)-1,
\eea
where $X=\Lambda,V$.  Using $\eta^{X}_T$, we now define the amplification of the incoherently scattered part of the probe beam as:
\bea
\eta^{X}_{\rm inc}=\eta^{X}_T-\eta^{X}_c. \label{Iamp}
\eea
For a $\Lambda$-type 3LE, we get from Eq.~\ref{Iamp}:
\bea
\eta^{\Lambda}_{\rm inc}=\f{2\Gamma_p}{v_gI_{\rm p}}(\mathcal{N}_2-|\mathcal{M}_1|^2).\label{IampL}
\eea
We derive the following relatively simple formula of $\eta^{\Lambda}_{\rm inc}$ by approximating the occupation $\mathcal{N}_2$ of the level $|3\ra$ and dropping the transition amplitude $\mathcal{M}_1$ between the levels $|2\ra$ and $|3\ra$ for a strong drive beam and a weak probe beam $(\Omega_d/\Omega_p \gg 1)$:
\bea
\eta^{\Lambda}_{\rm inc} \approx \f{2\Gamma_p\Gamma_{\gamma}\Omega_d^2}{v_gI_{\rm p}(\Gamma_{\gamma}(\tilde{\Gamma}^2+\Delta_d^2)+2\Omega_d^2(\Gamma_p+\Gamma_{\gamma}))}.\label{IampLa}
\eea
The above incoherent amplification of the probe beam is due to spontaneous emission from the level $|3\ra$, which is excited by the strong drive beam. Similarly, we can find for a $V$-type 3LE:
\bea
\eta^{V}_{\rm inc}=\f{2\Gamma_p}{v_gI_{\rm p}}(1-\mathcal{N}_4-\mathcal{S}_4-|\mathcal{S}_3|^2). \label{IampV}
\eea
For a strong drive beam and a weak probe beam $(\Omega_d/\Omega_p \gg 1)$, we approximate the occupation $(1-\mathcal{N}_4-\mathcal{S}_4)$ of the level $|2\ra$ and neglect the transition amplitude $\mathcal{M}_1$ between the levels $|1\ra$ and $|2\ra$ in Eq.~\ref{IampV}, and we find a simplified expression for $\eta^{V}_{\rm inc}$:
\bea
\eta^{V}_{\rm inc} \approx \f{2\Gamma_p\Gamma_{\gamma}\Omega_d^2}{v_gI_{\rm p}(2\Gamma_{p}((\Gamma_{\gamma}+2\Gamma_d)^2+\Delta_d^2)+\Omega_d^2(4\Gamma_p+\Gamma_{\gamma}))}.\nn\\\label{IampVa}
\eea
We find from the above approximate formulas in Eqs.~\ref{IampLa},\ref{IampVa} that the dependence of $\eta^{X}_{\rm inc}$ on the drive beam amplitude $\Omega_d$ is similar for $\Lambda$ and $V$-type system. While $\eta^{X}_{\rm inc}$ increases quadratically with $\Omega_d$ at low $\Omega_d$, it saturates at large $\Omega_d$. We also notice from these approximate formulas that $\eta^{X}_{\rm inc}$ increases linearly with $\Gamma_{\gamma}$ at small $\Gamma_{\gamma}$. While $\eta^{\Lambda}_{\rm inc}$ saturates quickly to some finite value at a higher $\Gamma_{\gamma}$,  $\eta^{V}_{\rm inc}$ falls at large $\Gamma_{\gamma}$ (which is not shown in Fig.~\ref{incohA}(d)). We plot these approximate formulas of $\eta^{X}_{\rm inc}$ in Fig.~\ref{incohA} both for $\Lambda$ and $V$-type 3LE by varying $\Omega_d$ and $\Gamma_{\gamma}$. We also include the exact dependence of  $\eta^{X}_{\rm inc}$ on  $\Omega_d$ and $\Gamma_{\gamma}$ in Fig.~\ref{incohA}. The approximate formulas in Eqs.~\ref{IampLa},\ref{IampVa} show an excellent match with the exact ones at low probe power for any $\Omega_d$ and $\Gamma_{\gamma}$. However, the agreement is not good at small $\Omega_d$ or $\Gamma_{\gamma}$ for high probe power (especially for $V$-type 3LE).

The spontaneous emission from the excited level $|3\ra$ to level $|2\ra$ (level $|2\ra$ to level $|1\ra$) is the primary source of incoherent amplification for $\Lambda$ ($V$). Such emission increases with increasing $\Omega_d$ before saturation of level $|3\ra$ at certain values of $\Omega_d$, and the emission is not much affected by the probe power \footnote{For a low drive power, the amount of spontaneous emission at the optical transition driven by the probe beam depends on the probe power.}. Therefore, we expect $\eta^{X}_{\rm inc}$ to increase with $\Omega_d$ before saturation and to fall with increasing $\Omega_p$, as shown in Fig.~\ref{incohA}. 
An increasing $\Gamma_{\gamma}$ can decrease the coherence of level $|2\ra$ of a $\Lambda$-type 3LE, but it improves the population inversion between levels $|2\ra$ and  $|3\ra$. Thus, the amplification of incoherently scattered probe beam increases with an increasing $\Gamma_{\gamma}$ before saturation. On the other hand, the population of level $|2\ra$ of a $V$-type 3LE first grows with an increasing $\Gamma_{\gamma}$, and then falls at larger $\Gamma_{\gamma}$; therefore, $\eta^{V}_{\rm inc}$ also increases with an increasing $\Gamma_{\gamma}$ before falling at large $\Gamma_{\gamma}$.

\subsection{Coherent vs. incoherent amplification}

\begin{figure}
\includegraphics[width=\columnwidth]{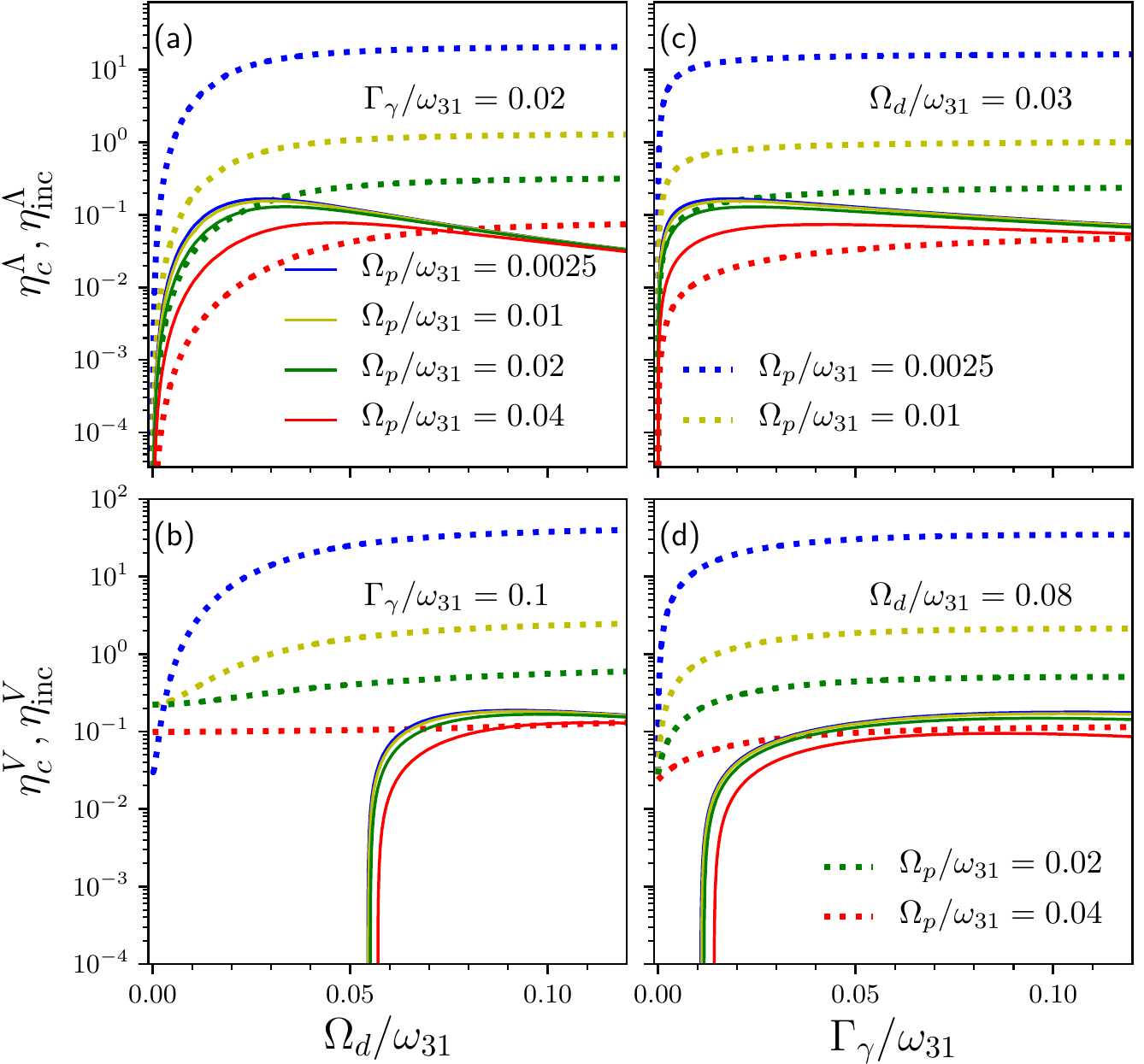}
\caption{Comparison between coherent amplification $\eta^{\Lambda}_c, \eta^{V}_c$ and incoherent amplification $\eta^{\Lambda}_{\rm inc}, \eta^{V}_{\rm inc}$ of a probe beam by respectively a $\Lambda$ and a $V$-type 3LE coupled to a 1D waveguide. (a,b) Scaling of $\eta^{\Lambda}_c, \eta^{V}_c,\eta^{\Lambda}_{\rm inc},\eta^{V}_{\rm inc}$ as a function of the Rabi frequency $\Omega_d$ of the drive beam, and (b,d) dependence of $\eta^{\Lambda}_c, \eta^{V}_c,\eta^{\Lambda}_{\rm inc},\eta^{V}_{\rm inc}$ on non-radiative decay $\Gamma_{\gamma}$. The dotted curves are for $\eta^{\Lambda}_{\rm inc}, \eta^{V}_{\rm inc}$ and the full lines are for $\eta^{\Lambda}_c, \eta^{V}_c$. The parameters are $\Gamma_p=\Gamma_d=0.01$ in all panels, and (a) $\Gamma_{\gamma}=0.02$, (c) $\Gamma_{\gamma}=0.1$, (b) $\Om_d=0.03$, (d) $\Om_d=0.08$. The parameters are in unit of $\om_{31}$.}
\label{cincohA}
\end{figure}

While coherent amplification has been mostly investigated for practical applications of these on-chip quantum amplifiers, incoherent amplification is also an integral part of such devices. The coherent amplification is mostly due to stimulated emission from the population-inverted emitter. The source of incoherent amplification is spontaneous emission at that transition. It is crucial to classify the parameter regimes, where the coherent and incoherent amplification dominate. For this, we here include a comparison between the efficiency of coherent and incoherent amplification in a $\Lambda$ and a $V$-type 3LE. In Fig.~\ref{cincohA}, we plot the exact lineshapes of $\eta^{X}_{c}$ and $\eta^{X}_{\rm inc}$ as a function of $\Omega_d$ and $\Gamma_{\gamma}$ for different probe power. We find $\eta^{X}_{\rm inc}$ much higher than $\eta^{X}_{c}$ for a weak probe beam in both a $\Lambda$ and a $V$-type 3LE at all $\Omega_d$. We further notice that  $\eta^{X}_{c}$ can be higher than $\eta^{X}_{\rm inc}$ at a higher probe power and a relatively smaller drive power where an on-chip quantum amplifier acts as a better coherent amplifier if we consider higher $\eta^{X}_{c}/\eta^{X}_{\rm inc}$ as efficiency criterion. However, we should remind that the maximum value of $\eta^{X}_{c}$ (as well as $\eta^{X}_{\rm inc}$) is obtained for a lower probe power. We also observe that both $\eta^{X}_{c}$ and $\eta^{X}_{\rm inc}$ themselves fall with increasing probe beam power at low drive power.

\subsection{Statistics of transmitted probe photons}

\begin{figure}
\includegraphics[width=\columnwidth]{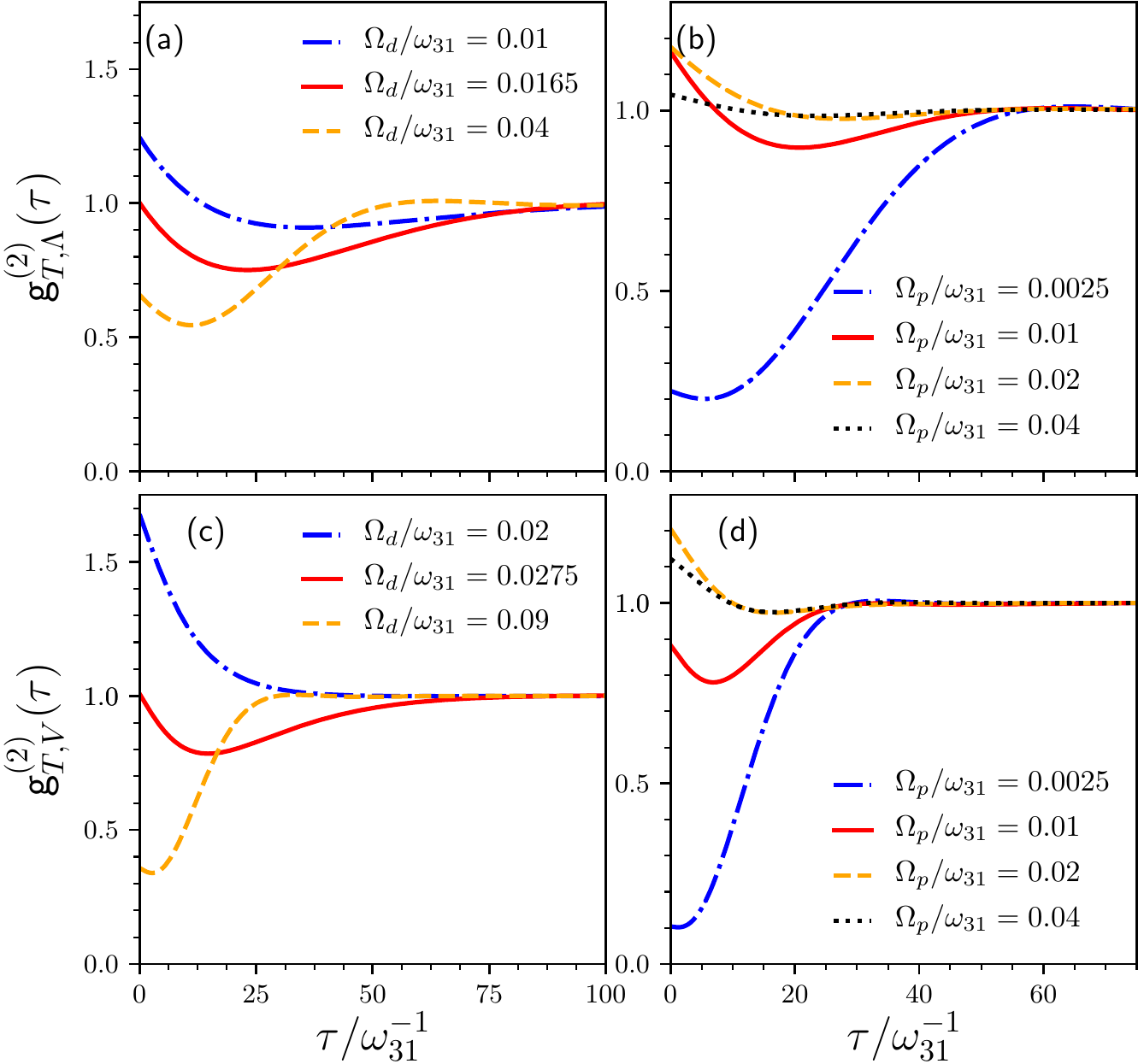}
\caption{Second-order correlation function $g^{(2)}_{T,\Lambda}(\tau), g^{(2)}_{T,V}(\tau)$  of a transmitted probe beam with time delay $\tau$ for different strength of the drive beam (panels a,c) and the probe beam (panels b,d). The parameters are $\Gamma_p=\Gamma_d=0.01,\Delta_p=\Delta_d=0$ in all panels, and (a,b) $\Gamma_{\gamma}=0.02$, (c,d) $\Gamma_{\gamma}=0.1$, (a,c) $\Omega_p=0.005$, (b) $\Omega_d=0.04$, (d) $\Omega_d=0.09$. The parameters are in the unit of $\omega_{31}$.}
\label{statA}
\end{figure}

Photon statistics is a crucial ingredient to study the physical nature of light, e.g., classical vs. quantum light. We here calculate the photon statistics of the transmitted probe light, which is amplified in the quantum amplifier modeled by a driven 3LE. We are particularly interested in understanding how amplification affects the statistics of transmitted probe photons. In our study, the initial states are coherent states which have Poissonian photon distribution. One standard measure of photon statistics is second-order (intensity) correlation function $g^{(2)}(\tau)$, which for the transmitted probe photon is defined as
\bea
g^{(2)}_{T,X}(\tau)=\f{\la a_{x+}^\dg(\tau)a_{x+}^\dg(t+\tau)a_{x+}(t+\tau)a_{x+}(\tau)\ra}{\langle a_{x+}^\dg(t)a_{x+}(t)\rangle\langle a_{x+}^\dg(t+\tau)a_{x+}(t+\tau)\rangle},
\eea
where $\tau$ is time delay, and the photon field is given by 
\bea
a_{x+}(t)&=&\zeta_a(t-\f{x}{v_g})-\f{i\sqrt{2\pi}g_p}{v_g}\zeta_X(t-\f{x}{v_g}), \\
\zeta_a(t)&=&\f{1}{\sqrt{2\pi}}\int_{-\infty}^{\infty}dk\:e^{-iv_gk(t-t_0)}a_{k+}(t_0),
\eea
where $\zeta_{\Lambda}=\mu$ and $\zeta_V=\sigma$. It can be shown that $\zeta_a(t+\tau-\f{x}{v_g})$ commutes with $\zeta_X(t-\f{x}{v_g})$ because our initial state is a product of the states of the 3LE and the photon fields. Such commutation simplifies the calculation of $g^{(2)}_{T,X}(\tau)$. 
By integrating out the photon fields after taking expectation over the initial photon fields, we can rewrite $g^{(2)}_{T,X}(\tau)$ as the following:
\bea
g^{(2)}_{T,X}(\tau)=\f{G_T^X(t,\tau)}{\mathcal{T}^X_p(t)\mathcal{T}^X_p(t+\tau)},~{\rm where}\label{2ndcoh}
\eea
\begin{widetext}
\bea
&&G_T^X(t,\tau)\nn\\&=& \mathcal{T}^X_p(t+\tau)+\mathcal{T}^X_p(t)-1-\f{16\Gamma_p^3}{\Omega_p^3}{\rm Im}[\la\zeta^\dg_X(t')\zeta^\dg_X(t'+\tau)\zeta_X(t')\ra e^{-i\om_p(t'+\tau-t_0)}+\la\zeta^\dg_X(t')\zeta^\dg_X(t'+\tau)\zeta_X(t'+\tau)\ra e^{-i\om_p(t'-t_0)}]\nn\\
&+&\f{8\Gamma_p^2}{\Omega_p^2}{\rm Re}[\la\zeta^\dg_X(t')\zeta_X(t'+\tau)\ra e^{i\om_p\tau}-\la\zeta^\dg_X(t')\zeta^\dg_X(t'+\tau)\ra e^{-i\om_p(2t'+\tau-2t_0)}]+\f{16\Gamma_p^4}{\Omega_p^4}\la\zeta^\dg_X(t')\zeta^\dg_X(t'+\tau)\zeta_X(t'+\tau)\zeta_X(t')\ra, \label{correlations}
\eea
\end{widetext}
with $t'=t-\f{x}{v_g}$. The second-order correlation $g^{(2)}_{T,X}(\tau)=1$ for a coherent state with a Poissonian distribution of photons.  While $g^{(2)}_{T,X}(\tau=0)>1$ indicates photon bunching and super-Poissonian distribution of photons, the light is anti-bunched and has sub-Poissonian distribution of photons when $g^{(2)}_{T,X}(\tau=0)=0$. We can obtain a simple form for $g^{(2)}_{T,X}(\tau=0)$ by setting $\tau=0$ in $G_T^X(t,\tau)$ in Eq.~\ref{correlations}, and we find
\bea
g^{(2)}_{T,X}(0)= \f{2( \mathcal{R}^{X}_{p}+\mathcal{T}^{X}_{p})-1}{(\mathcal{T}^{X}_{p})^2}. \label{2ndcoh0}
 \eea
 In the absence of non-radiative decay, there is no exchange of photons between the probe and drive beams, and we have $\mathcal{R}^{X}_{p}+\mathcal{T}^{X}_{p}=1$. Therefore, we then get $g^{(2)}_{T,X}(0)= (\mathcal{T}^{X}_{p})^{-2} \ge 1$ (equality for $\mathcal{R}^{X}_{p}=0$) signaling bunching of transmitted probe photons in both $\Lambda$ and $V$-type systems in the absence of amplification of the probe beam and non-zero reflection. We find from Eq.~\ref{2ndcoh0} that $g^{(2)}_{T,X}(0)=1$ when $2\mathcal{R}^{X}_{p}=(\mathcal{T}^{X}_{p}-1)^2$, which is feasible  for a non-zero $\mathcal{R}^{X}_{p}$ only in the presence of amplification of the probe beam. We can further argue  from Eq.~\ref{2ndcoh0} by multiplying the numerator and denominator by $I_{\rm p}^2$ that $g^{(2)}_{T,X}(0) \to 0$ (anti-bunching) when probe intensity $I_{\rm p} \to 0$ in the presence of amplification.
 
The two-time correlators in Eq.~\ref{correlations} can be derived using a set of differential equations of the form of Eq.~\ref{eom3LEq}. Let us define a set of operators $\boldsymbol{\mathcal{V}}_X$ such that $\la\boldsymbol{\mathcal{V}}_X\ra=\boldsymbol{\mathcal{M}}_X$, and we have from Eqs.~\ref{eom3LEq}, \ref{eom4LEq}:
\bea
&&\f{d\la\boldsymbol{\mathcal{V}}_X\ra}{dt}=\boldsymbol{\mathcal{R}}_{X}\la\boldsymbol{\mathcal{V}}_X\ra+\boldsymbol{\Omega}_{X}.
\eea
Due to the Markovian dynamics of our current waveguide QED systems and the product form of the initial state, we get the following differential equations for the required two-time correlators of the operators using the quantum regression theorem \cite{RoyPRA2017}:
\bea
\f{d\la\zeta^\dg_X(t')\boldsymbol{\mathcal{V}}_X(t'+\tau)\ra}{d\tau}&=&\boldsymbol{\mathcal{R}}_{X}\la\zeta^\dg_X(t')\boldsymbol{\mathcal{V}}_X(t'+\tau)\ra\nn\\ &&+\boldsymbol{\Omega}_{X}\la\zeta^\dg_X(t')\ra,\\
\f{d\la\zeta^\dg_X(t')\boldsymbol{\mathcal{V}}_X(t'+\tau)\zeta_X(t')\ra}{d\tau}&=&\boldsymbol{\mathcal{R}}_{X}\la\zeta^\dg_X(t')\boldsymbol{\mathcal{V}}_X(t'+\tau)\zeta_X(t')\ra \nn\\ &&+\boldsymbol{\Omega}_{X}\la\zeta^\dg_X(t')\zeta_X(t')\ra, \\
\f{d\la\boldsymbol{\mathcal{V}}_X(t'+\tau)\zeta_X(t')\ra}{d\tau}&=&\boldsymbol{\mathcal{R}}_{X}\la\boldsymbol{\mathcal{V}}_X(t'+\tau)\zeta_X(t')\ra\nn\\ &&+\boldsymbol{\Omega}_{X}\la\zeta_X(t')\ra.
\eea
We solve these equations in steady-state for the correlators and use them in Eq.~\ref{2ndcoh} to calculate the coherence properties of the amplified photons. In Fig.~\ref{statA}, we show $g^{(2)}_{T,X}(\tau)$ with delay time $\tau$ of the transmitted probe beam from a $\Lambda$ and a $V$-type 3LE for different strength of the drive and probe beams. For a weak probe beam, we expect $g^{(2)}_{T,X}(0)\ge 1$ for lower amplification at weak driving, and $g^{(2)}_{T,X}(0)<1$ for higher amplification at stronger driving, as discussed above. We display these features in Fig.~\ref{statA}(a,c) for a $\Lambda$ and a $V$-type 3LE, respectively. In Fig.~\ref{statA}(b,d), we further plot $g^{(2)}_{T,X}(\tau)$ for increasing probe beam power at a constant drive beam strength. We find $g^{(2)}_{T,X}(0)$ can become nearly zero for low probe beam power where the incoherent amplification dominates. At this probe power regime, the transmitted probe beam mostly consists of spontaneously emitted photons from the drive-beam excited emitter, and the feature of $g^{(2)}_{T,X}(0)$ is determined predominantly by these spontaneously emitted photons. On the other hand, the value of $g^{(2)}_{T,X}(0)$ first increases with increasing probe power and then decreases to 1 for further increasing probe power when the 3LE is saturated.   

\section{Cross-Kerr nonlinearity}
\label{cross}
An effective interaction between different light fields at the single-photon quantum regime is essential for many all-optical quantum devices and quantum logic gates \cite{Brod2016, Liu2016, Zhang2017}. The waveguide QED systems are regarded to be perfectly suitable for creating such interaction between propagating photons by cross-Kerr coupling in a nonlinear medium of single or multiple emitters. A large cross-Kerr phase shift per photon has been demonstrated with two coherent microwave fields at a single-photon level in a transmission line strongly coupled to a  ladder-type 3LE made of superconducting artificial atom \cite{HoiPRL2013}. Such effective photon-photon interaction in the cross-Kerr medium has been used to propose quantum nondemolition measurement of a single propagating microwave photon with high fidelity \cite{Sathyamoorthy2014}. Cross-Kerr nonlinearity is also often employed in various schemes of generation of entanglement between photons \cite{Xiu2016, Wang2015}. Therefore, it is an important question to find out which type of 3LE creates stronger effective photon-photon interaction in waveguide QED.

The cross-Kerr effect in bulk media is usually interpreted as modulation of the refractive index due to the application of a strong drive field. For many materials, the refractive index $n$ of the probe beam in the presence of a drive beam can be written \cite{Boyd08} as
\bea
n=n_0+\bar{n}^{\rm cross}_2|E_d|^2,
\eea   
where $n_0$ is the weak-drive refractive index, and $\bar{n}^{\rm cross}_2$ is a nonlinear coefficient which is proportional to the third-order susceptibility $\chi^{(3)}$. Therefore, a change in the refractive index is given by
\bea
\Delta n=\bar{n}^{\rm cross}_2|E_d|^2,\label{crossS}
\eea
which captures the cross-coupling between the probe and drive beams. For light scattering by a single emitter inside the waveguide, we can connect the complex susceptibility of the material to the change in phase of coherently scattered probe photons. The nonlinear response due to coherent interaction between the drive and probe fields at the emitter then is related to the difference of such change in phase $\Delta\phi_p^X$ of the probe field in the presence and absence of the drive field. In the regime of experimental interest with a single probe and drive photons \cite{HoiPRL2013}, we can express the above relation in Eq.~\ref{crossS} in the following approximate form:
\bea
\Delta\phi_p^X=k_{X}\Om_d^2,
\eea
where $k_{X}$ is the Kerr coefficient which we use in the following.

We here make a detailed theoretical analysis of cross-Kerr nonlinearity mediated by a $\Lambda$ or a $V$ or a ladder-type 3LE embedded in an open 1D waveguide.We make a comparison between cross-Kerr phase shifts from these three systems to quantify their performances. We also relate the phase response to the amplitude response of the probe beam in these systems, which would eventually connect the cross-Kerr phase shift and the coherent amplification. In the following discussion of cross-Kerr effect, we take non-zero pure-dephasing and set non-radiative decay $\Gamma_{\gamma}=0$, which implies these systems do not amplify the probe beam. We later include non-radiative decay to explore a connection between the phase and the amplitude response.

\subsection{Comparison of cross-Kerr phase shift from $\Lambda$, $V$, and ladder-type 3LE}

\begin{figure*}
\includegraphics[width=\linewidth]{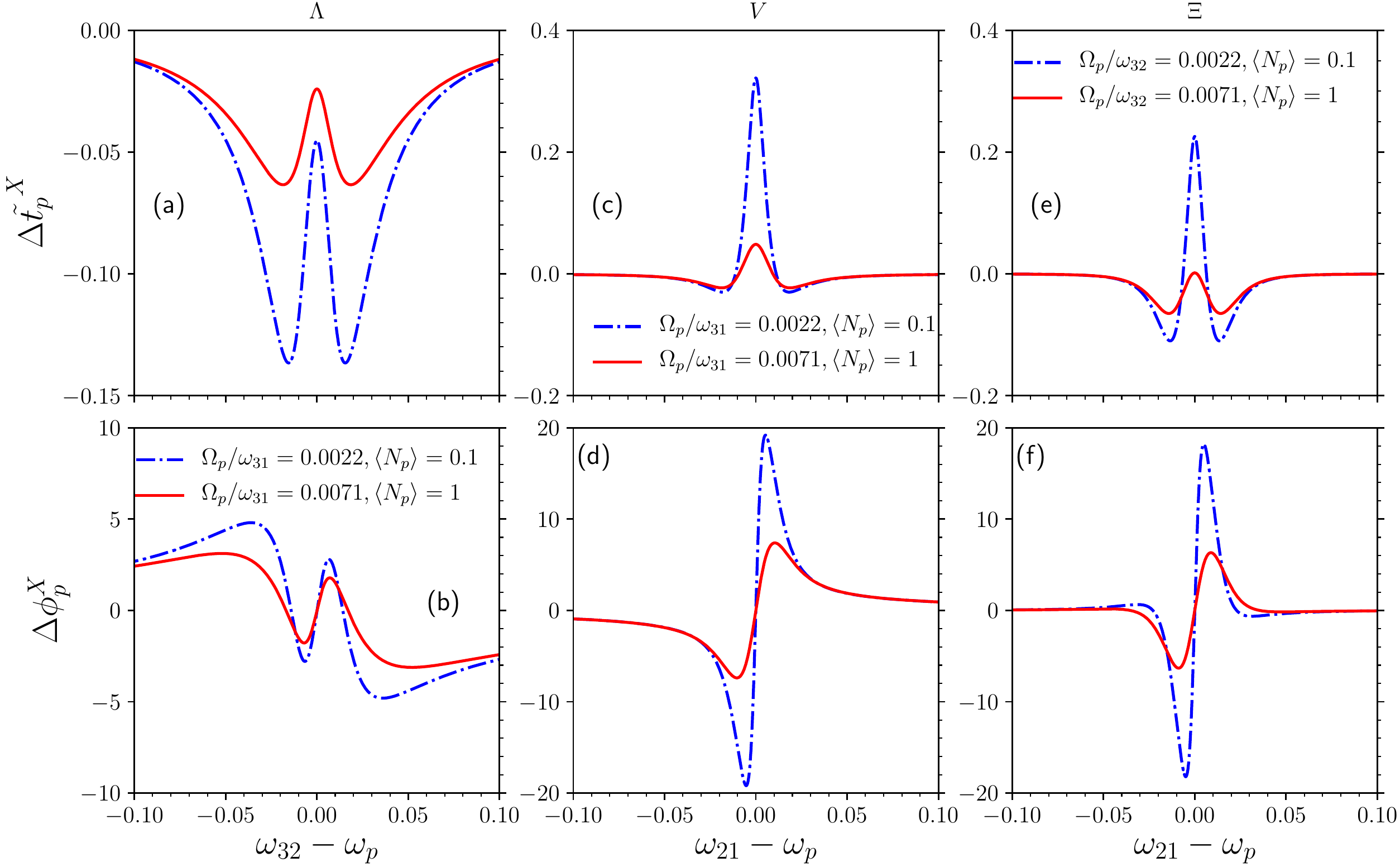}
\caption{Amplitude and phase response ($\Delta \tilde{t}^X_p$ and $\Delta \phi^X_p$) of probe transmission through a $\Lambda$, $V$, and $\Xi$-type 3LE embedded in a 1D waveguide as a function of probe beam detuning. The parameters are $\Gamma_d=0.005,\Gamma_p=0.0025,\Gamma_{\lambda 2}=0.0028,\Gamma_{\lambda 3}=0.0118, \Gamma_{\gamma}=0, \Delta_d=0, \Omega_d=0.01414~(\la N_d\ra=1)$. The parameters are in unit of $\omega_{31}$ for $\Lambda$ and $V$-type 3LE and in unit of $\omega_{32}$ for $\Xi$-type 3LE.}
\label{APresponse}
\end{figure*}

\textcite{HoiPRL2013} have quantified the cross-Kerr nonlinearity through a difference in phase of the coherent transmission amplitude of the probe field in the presence ($\Omega_d\ne 0$) and absence ($\Omega_d=0$) of the drive beam. The coherent transmission amplitude ${\tilde t}_p^X$ of the probe beam incoming from the left of the $X$-type 3LE can be defined as
\bea
    {\tilde{t}}_p^X&=&\f{\la \psi| a^{\dg}_{x>0,+}(t)|\psi\ra}{\la \psi| a^{\dg}_{x>0,+}(t)|\psi\ra_{g_p=0}}=1+2i\boldsymbol{\chi}^X(t,\Delta_p,\Delta_d) \nn\\
    &=&1+\f{2i\Gamma_p}{\Omega_p}e^{-i\omega_p(t-x/v_g-t_0)}\la\zeta^{\dg}_X(t-\f{x}{v_g})\ra,
\eea
where $\zeta_{\Lambda}=\mu,\zeta_{V,\Xi}=\sigma$, and $\la \dots \ra_{g_p=0}$ indicates no coupling between the emitter and the incident probe field. Here, $\boldsymbol{\chi}^X(t,\Delta_p,\Delta_d)$ represents the optical susceptibility of the medium, which includes both linear and nonlinear parts of the susceptibility. At long-time steady-state, $\boldsymbol{\chi}^X(t,\Delta_p,\Delta_d)$ becomes independent of time, and we here onward consider the cross-Kerr nonlinearity at steady-state. The phase $\phi_{p}^X$ associated with ${\tilde t}_p^X$ at steady-state (by dropping the time variable in $\boldsymbol{\chi}^X(t,\Delta_p,\Delta_d)$) is given by
\bea
\phi_{p}^X=\tan^{-1}{\Big(\f{2\:{\rm Re} \boldsymbol{\chi}^X(\Delta_p,\Delta_d)}{1-2\:{\rm Im} \boldsymbol{\chi}^X(\Delta_p,\Delta_d)}\Big)}. \label{presponse}
\eea
Therefore, the cross-Kerr phase shift $\Delta\phi_{p}^X=\phi_{p}^X|_{\Omega_d\ne 0}-\phi_{p}^X|_{\Omega_d=0}$. Similarly, we define the amplitude response $\Delta \tilde{t}^X_p$ of the probe beam as a difference between the magnitude of the probe transmission amplitude in the presence ($\Omega_d\ne 0$) and absence ($\Omega_d=0$) of the drive beam, $\Delta\tilde{t}_{p}^X=|\tilde{t}_{p}^X|_{\Omega_d\ne 0}-|\tilde{t}_{p}^X|_{\Omega_d=0}$. In Fig.~\ref{APresponse}, we plot the amplitude and phase response  $\Delta\tilde{t}_{p}^X, \Delta\phi_{p}^X$ of probe transmission as a function of detuning of the probe beam from the related optical transition of an $X$-type 3LE. We take smaller values of probe and drive power to examine the responses of probe transmission at quantum regime with relatively low incoherent scattering. Following Ref.~\cite{HoiPRL2013}, we define the average number of probe (drive) photons $\la N_p\ra$ ($\la N_p\ra$) per interaction time, $1/4\Gamma_p$ ($1/4\Gamma_d$) as $\la N_p\ra=\Omega_p^2/8\Gamma_p^2$ ($\la N_d\ra=\Omega_d^2/8\Gamma_d^2)$. We show probe responses in Fig.~\ref{APresponse} for two different values of  $\la N_p\ra=0.1,1$ and a small $\la N_d\ra=1$. We choose the values of $\Gamma_p,\Gamma_d,\Gamma_{\lambda_1}$ and $\Gamma_{\lambda_2}$ to be similar to those in Ref.~~\cite{HoiPRL2013} for a ladder system.   

While $V$ and $\Xi$ systems act as a 2LE for a probe beam in the absence of a drive beam, the probe beam ceases to interact with a $\Lambda$ system as the drive field is turned off. Therefore, $|\tilde{t}_{p}^X|_{\Omega_d=0}$ is that of a 2LE for $X=V,\Xi$ (depicting perfect reflection or zero transmission at zero detunings), and $\tilde{t}_{p}^{\Lambda}|_{\Omega_d=0}=1$. Thus, $\Delta \tilde{t}^{\Lambda}_p$ in Fig.~\ref{APresponse}(a) depicts the transmission amplitude (shifted downwards by 1) of a probe beam manifesting a peak at zero probe detuning (the detuning of the drive beam is fixed to zero) due to electromagnetically induced transparency in the presence of a drive beam. We also find the peak height at two-photon resonance increases with increasing  $\la N_p\ra$ \cite{RoyPRL2011}. The probe beam transmits through the system without interacting with the emitter at large probe detuning, and  $\Delta \tilde{t}^{\Lambda}_p$ becomes nearly zero. The interaction of a probe beam with a side-coupled 2LE, $V$ and $\Xi$ system decreases with increasing  probe detuning. Therefore, the transmission becomes close to one. Nevertheless, $\Delta \tilde{t}^{V,\Xi}_p$ is almost zero at large probe detuning in Fig.~\ref{APresponse}(c,e) due to a difference between two numbers, which are nearly equal to one.  $|\tilde{t}_{p}^{V,\Xi}|_{\Omega_d=0}$ depicts perfect reflection or zero transmission at zero detunings, and the drive beam again induces transparency in $|\tilde{t}_{p}^{V,\Xi}|_{\Omega_d \ne 0}$ near two-photon resonance \cite{Witthaut10}. The peak height of $\Delta\tilde{t}_{p}^{V,\Xi}$ at zero probe detuning decreases with increasing  $\la N_p\ra$ due to saturation of the 3LE by the probe beam in the presence and absence of the drive beam.

In the bottom row of Fig.~\ref{APresponse}, we plot the phase responses $\Delta\phi_{p}^X$ of probe transmission corresponding to those amplitude responses in the top row of Fig.~\ref{APresponse}. The main observations on the phase responses in all three systems are the following: (i) the extrema of $\Delta\phi_{p}^X$ appear at some finite detuning of the probe beam depending on $\la N_p\ra$, (ii) the magnitude of extrema of $\Delta\phi_{p}^X$ decreases with increasing $\la N_p\ra$ due to saturation of the emitter by the beams, (iii) the position of the extrema of $\Delta\phi_{p}^X$ in probe detuning lies in between the extrema of the amplitude response where the amplitude response changes rapidly. We also observe from  Fig.~\ref{APresponse} (b,d,f) that the maximum value of $\Delta\phi_{p}^X$ is relatively high for $V$ and $\Xi$ systems in comparison to the $\Lambda$ system at these probe and drive power.  While the inclusion of non-radiative decay causes a decrease of cross-Kerr phase shift in a $\Lambda$ and a ladder system, it can improve or deteriorate the cross-Kerr phase shift in a $V$ system depending on the parameters.

\begin{figure}
\includegraphics[width=\linewidth]{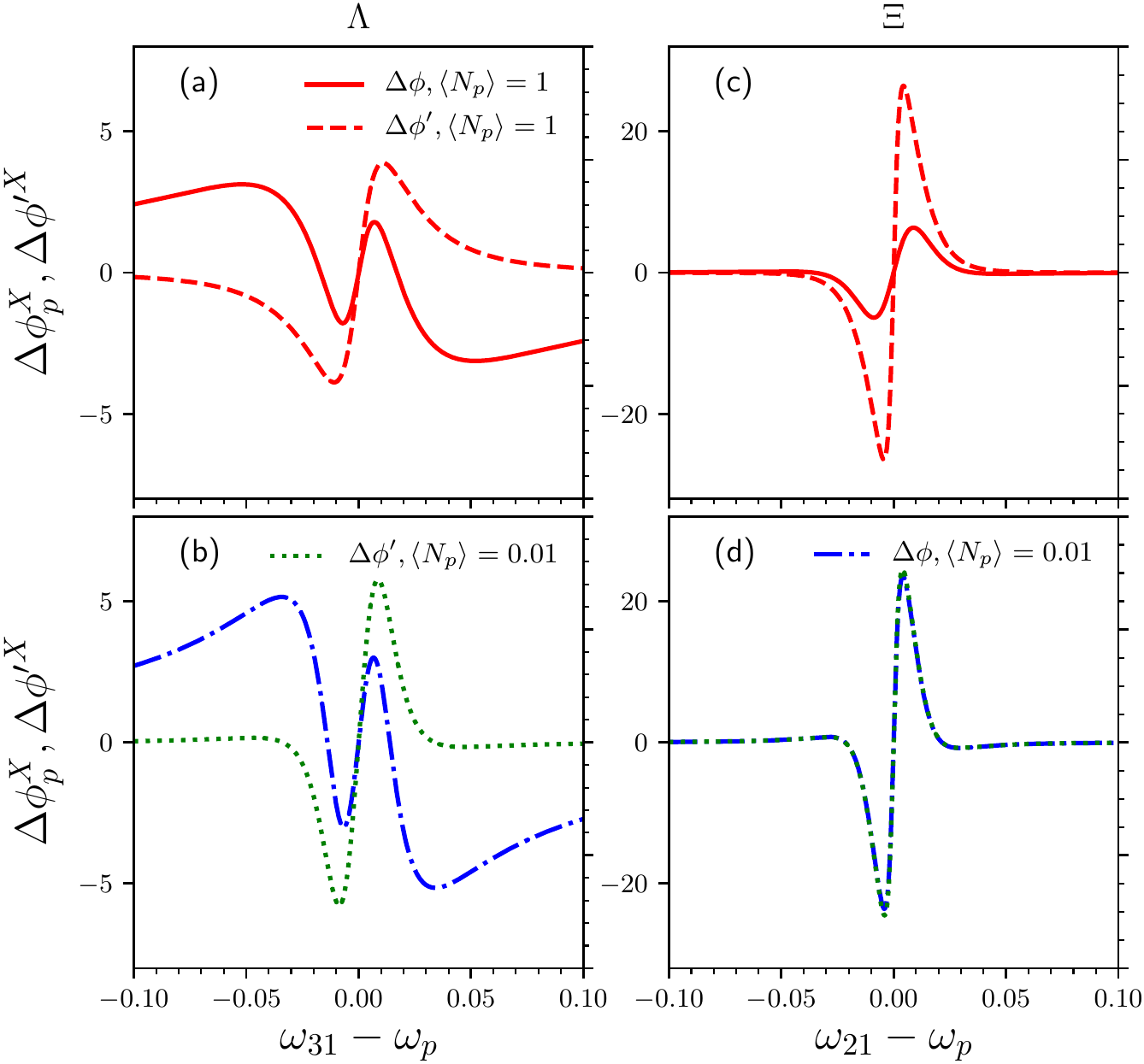}
\caption{Comparison between different definitions of phase response ($\Delta \phi^X_p$ and $\Delta \phi^{'X}_p$) of probe transmission through a $\Lambda$ and $\Xi$-type 3LE as a function of the probe beam detuning. The parameters are $\Gamma_d=0.005,\Gamma_p=0.0025,\Gamma_{\lambda 2}=0.0028,\Gamma_{\lambda 3}=0.0118, \Gamma_{\gamma}=0, \Delta_d=0, \Omega_d=0.01414~(\la N_d\ra=1)$. The parameters are in unit of $\omega_{31}$ for $\Lambda$-type 3LE and in unit of $\omega_{32}$ for $\Xi$-type 3LE.}
\label{Presponse}
\end{figure}

We find from Fig.~\ref{APresponse} that unlike $V$ and $\Xi$ systems, the phase response in a $\Lambda$ system does not rapidly reduce to zero with increasing probe detuning.  This is due to zero value of $\phi_{p}^{\Lambda}|_{\Omega_d=0}$, which results from no interaction of the probe beam with the $\Lambda$-type 3LE in the absence of the drive beam. Nevertheless, there is a linear response regime for a weak probe beam in the presence of a weak drive beam in a $\Lambda$-type 3LE, e.g., $\Omega_p,\Omega_d \ll \Gamma_p,\Gamma_d,\Gamma_{\lambda 2},\Gamma_{\lambda 3}$, where the probe transmission resembles that from a 2LE. In this linear response regime, we have an approximate form for $\boldsymbol{\chi}^{\Lambda}(\om_p,\om_d)$ as
\bea
\boldsymbol{\chi}^{\Lambda}_l=-\f{\Gamma_p}{\Delta_p+i\Gamma_{32}}\Big(1-\f{\Gamma_{31}\Gamma_{32}(I_{\rm p}/I_{\rm d})}{\Gamma_{32}^2+\Gamma_{31}\Gamma_{32}(I_{\rm p}/I_{\rm d})+\Delta_p^2}\Big),
\eea
with $\Gamma_{31}=\tilde{\Gamma}+\Gamma_{\lambda 3},\Gamma_{21}=\Gamma_{\lambda 2},\Gamma_{32}=\Gamma_{31}+\Gamma_{21}$. We apply $\boldsymbol{\chi}^{\Lambda}_l$ in Eq.~\ref{presponse} to calculate the phase $\phi^{'\Lambda}_p$ of the transmission amplitude of the probe field in the linear regime. Using $\phi^{'\Lambda}_p$, we propose a new definition of the phase response as $\Delta \phi^{'\Lambda}_p=\phi^{\Lambda}_p|_{\Omega_d\ne 0}-\phi^{'\Lambda}_p|_{\Omega_d\ne 0}$. In Fig.~\ref{Presponse}(a,b), we show the lineshape of $\Delta \phi^{'\Lambda}_p$ as a function of probe detuning and compare it with $\Delta \phi^{\Lambda}_p$. $\Delta \phi^{'\Lambda}_p$ vanishes at large probe detuning, and it also  gives higher magnitude for cross-Kerr phase shift than $\Delta \phi^{\Lambda}_p$. To further investigate the effectiveness of the new definition of the cross-Kerr phase shift, we also plot $\Delta \phi^{'\Xi}_p~(\equiv \phi^{\Xi}_p|_{\Omega_d\ne 0}-\phi^{'\Xi}_p|_{\Omega_d\ne 0})$ in Fig.~\ref{Presponse}(c,d) where we use the linear form, $\boldsymbol{\chi}^{\Xi}_l=-\Gamma_p/(\Delta_p+i(2\Gamma_p+\Gamma_{\lambda 2}))$ to find $\phi^{'\Xi}_p$ using Eq.~\ref{presponse}. While $\Delta \phi^{'\Xi}_p$ matches with $\Delta \phi^{\Xi}_p$ for a small probe power in Fig.~\ref{Presponse}(d), they differ quite a bit at a relatively higher probe power in Fig.~\ref{Presponse}(c). This is probably due to the absence of the self-Kerr effect of the probe beam in $\Delta \phi^{\Xi}_p$ against its  presence in $\Delta \phi^{'\Xi}_p$. Therefore, we conclude that $\Delta \phi^{'\Lambda}_p$ properly quantifies the cross-Kerr phase shift in the $\Lambda$ system only at relatively low probe power.

\begin{figure}
\includegraphics[width=\linewidth]{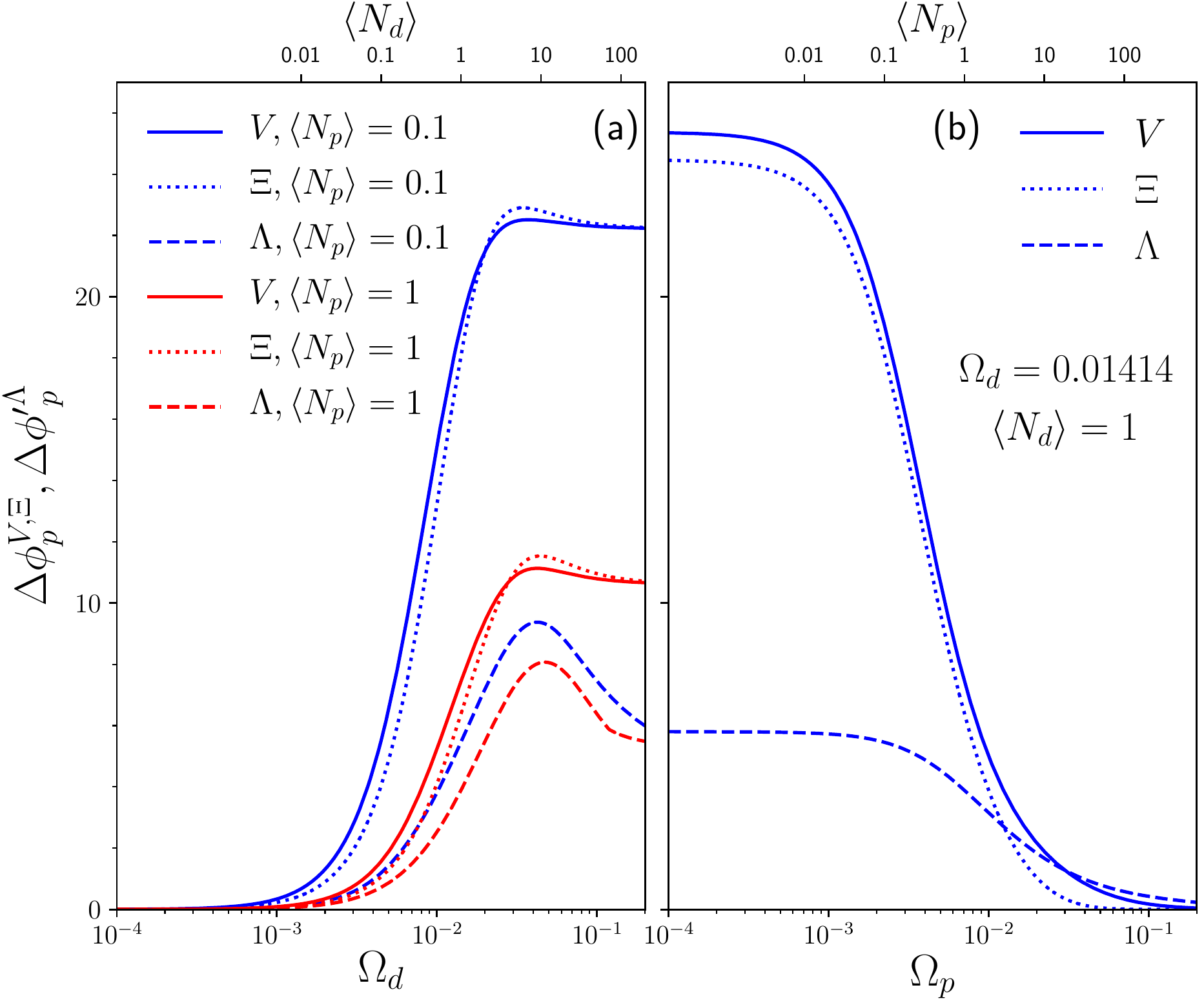}
\caption{Phase response ($\Delta \phi^{V,\Xi}_p$ and $\Delta \phi^{'\Lambda}_p$) as a function of the Rabi frequency $\Omega_p,\Omega_d$ (also $\la N_p\ra, \la N_d\ra$) of the probe and drive beams at a probe frequency that maximizes the phase shift. The parameters are $\Gamma_d=0.005,\Gamma_p=0.0025,\Gamma_{\lambda 2}=0.0028,\Gamma_{\lambda 3}=0.0118, \Gamma_{\gamma}=0, \Delta_d=0$. The parameters are in unit of $\omega_{31}$ for $\Lambda$-type 3LE and in unit of $\omega_{32}$ for $\Xi$-type 3LE.}
\label{Presponse1}
\end{figure}

Finally, we discuss the features of the phase response of probe transmission in different 3LEs as a function of the probe and drive beam power at a probe frequency that maximizes the phase shift. These features were measured in Ref.~\cite{HoiPRL2013} for a ladder system. In Fig.~\ref{Presponse1}(a), we plot $\Delta \phi^{V,\Xi}_p$ and $\Delta \phi^{'\Lambda}_p$ as a function of $\Omega_d$ and $\la N_d\ra$ for two different values of $\la N_p\ra$ at a probe frequency that maximizes the phase shift. We find that $\Delta \phi^{V,\Xi}_p$ increases with increasing $\Omega_d~(\la N_d\ra)$ before saturating at a relatively large $\la N_d\ra/\la N_p\ra$ values. We notice that the extremum of $\phi^{X}_p$ appears at a very different probe frequency for $\Omega_d=0$ and large $\Omega_d$. We also observe the magnitude of the extremum of $\phi^{X}_p$ decreases with increasing $\Omega_d$. Therefore, the saturation value of $\Delta \phi^{X}_p$ in Fig.~\ref{Presponse1}(a) at large $\Omega_d$ is mostly determined by the extremum of $\phi^{X}_p$ at $\Omega_d=0$. We further notice  from Fig.~\ref{Presponse1}(a) that the magnitude of $\Delta \phi^{X}_p$ decreases for a higher value of $\la N_p\ra$ at any $\Omega_d~(\la N_d\ra)$ which we depict in Fig.~\ref{Presponse1}(b) both for $\Delta \phi^{V,\Xi}_p$ and $\Delta \phi^{'\Lambda}_p$. While the constant cross-Kerr phase shift at very small $\Omega_p$ in  Fig.~\ref{Presponse1}(b) denotes the linear probe regime, the decrease of $\Delta \phi^{V,\Xi}_p$ and $\Delta \phi^{'\Lambda}_p$ with increasing $\Omega_p$ is due to the saturation of probe transition by the probe beam. The Fig.~\ref{Presponse1}(a) shows a strong non-monotonic dependence of $\Delta \phi^{'\Lambda}_p$ on $\Omega_d$ (or $\la N_d\ra)$ for a fixed $\la N_p\ra$. Such non-monotonic dependence is generated by competition between the phase shift near the small probe detuning at zero (or very small) $\Omega_d$ and that near the Autler-Townes peaks at higher $\Omega_d$.

At relatively small values of $\Omega_d~(\la N_d\ra)$, we find $\Delta \phi^{V}_p>\Delta \phi^{\Xi}_p\gg\Delta \phi^{'\Lambda}_p$, which can be argued by comparing the Kerr coefficient $k_X$ defined as $\Delta \phi^{V,\Xi}_p=k_{V,\Xi}\Omega_d^2$ or $\Delta \phi^{'\Lambda}_p=k_{\Lambda}\Omega_d^2$ \cite{HoiPRL2013}. In the parameter regime, $\Omega_p  \ll \Omega_d<\Gamma_p,\Gamma_d,\Gamma_{\lambda2},\Gamma_{\lambda3}$, we derive approximate $k_X$ for different 3LEs as
\bea
k_{\Lambda}&=&\f{2\Gamma_p\Delta_p(\Gamma_{21}\Gamma_{32}+(\Gamma_{21}+\Gamma_{32})(\Gamma_{32}-2\Gamma_p)-\Delta_p^2)}{(\Gamma_{21}^2+\Delta_p^2)(\Gamma_{32}^2+\Delta_p^2)((\Gamma_{32}-2\Gamma_p)^2+\Delta_p^2)},\nn\\
k_{V}&=&\f{\Gamma_p\Delta_p(\Gamma_{21}'^2+4\Gamma_d(\Gamma'_{21}-\Gamma_p)+\Delta_p^2)}{\Gamma_{31}'\Gamma_d(\Gamma_{21}'^2+\Delta_p^2)((\Gamma'_{21}-2\Gamma_p)^2+\Delta_p^2)},\nn\\
k_{\Xi}&=&\f{2\Gamma_p\Delta_p(\Gamma'_{21}\Gamma'_{31}+(\Gamma'_{21}+\Gamma'_{31})(\Gamma'_{21}-2\Gamma_p)-\Delta_p^2)}{(\Gamma_{21}'^2+\Delta_p^2)(\Gamma_{31}'^2+\Delta_p^2)((\Gamma'_{21}-2\Gamma_p)^2+\Delta_p^2)},\nn
\eea
where $\Gamma_{21}'=2\Gamma_p+\Gamma_{\lambda 2}, \Gamma_{31}'=2\Gamma_d+\Gamma_{\lambda 3},\Gamma_{32}'=\Gamma_{31}'+\Gamma_{21}'$. We find for a small cross-Kerr phase shift in the regime $\Omega_p\ll \Omega_d < \Gamma_d,\Gamma_p,\Gamma_{\lambda2},\Gamma_{\lambda3}$: $|k_{V}|>|k_{\Xi}|,|k_{\Lambda}|$, which shows a $V$ system can induce a higher cross-Kerr nonlinearity than a ladder or a $\Lambda$ system. We further find $|k_{\Xi}|>|k_{\Lambda}|$ implying better performance of a ladder system over a $\Lambda$ system in the above regime of small cross-Kerr phase shift.

\subsection{Amplitude and phase response: Kramers-Kronig relations}
\begin{figure}
\includegraphics[width=\linewidth]{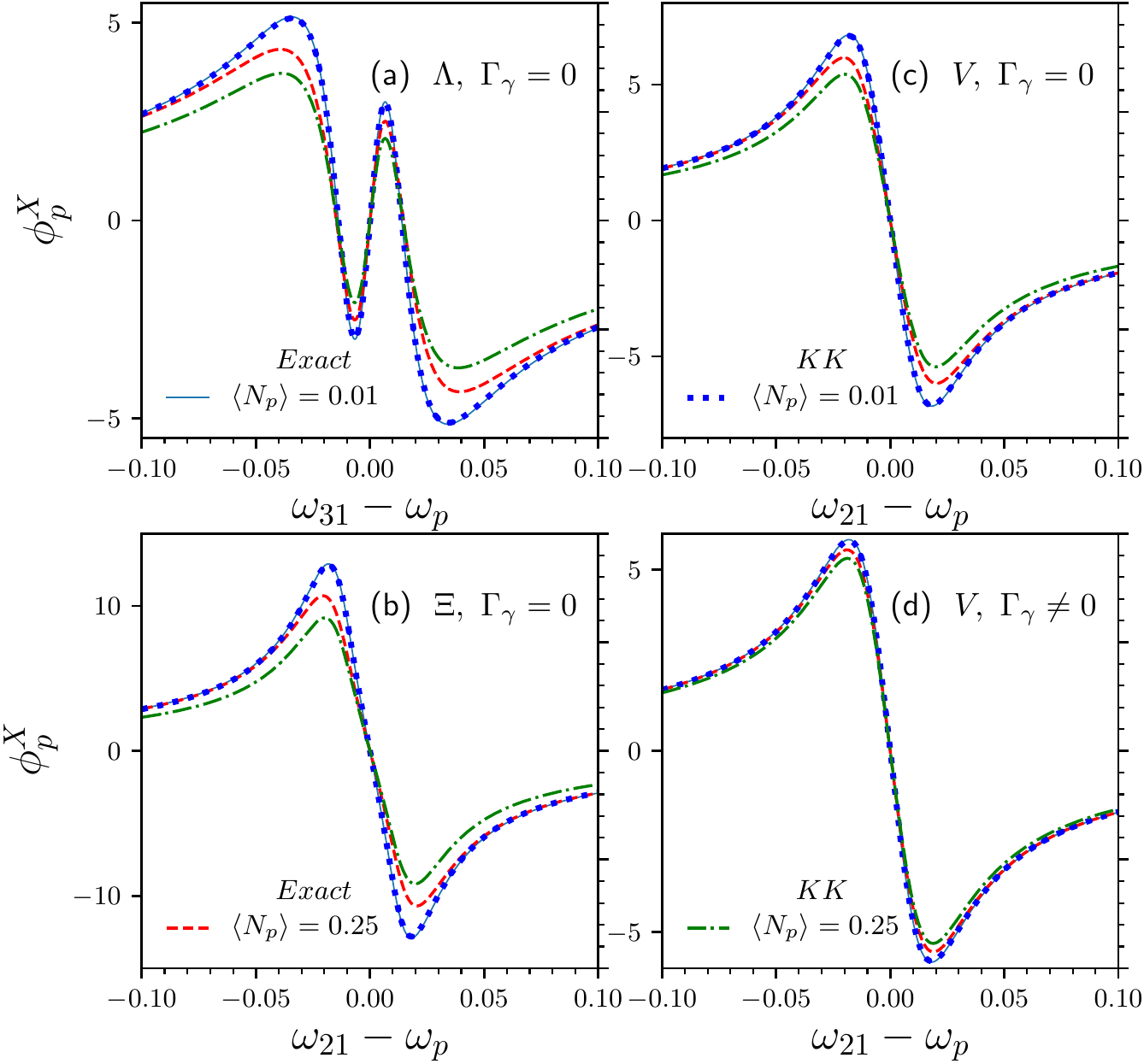}
\caption{Comparison between the exact phase response $\phi^{X}_p$ ($X=\Lambda,\Xi,V$), and that derived using the Kramers-Kronig (KK) relation. The parameters are $\Gamma_d=0.005,\Gamma_p=0.0025,\Gamma_{\lambda 2}=0.0028,\Gamma_{\lambda 3}=0.0118,\Gamma_{\gamma}=0.001, \Delta_d=0,\Omega_d=0.01414$. The parameters are in unit of $\omega_{31}$ for $\Lambda$-type 3LE and in unit of $\omega_{32}$ for $\Xi$-type 3LE.}
\label{kkfig}
\end{figure}

We have discussed coherent amplification and cross-Kerr phase shift of probe transmission using the amplitude and phase response of the transmitted probe field. We here show that these two responses are related by the well-known Kramers-Kronig relations which connect the real and imaginary parts of any complex function that is analytic in the upper half-plane of a complex variable and vanishes at a specific rate as the magnitude of the complex variable goes to infinity. We can write the probe transmission in steady-state as ${\tilde{t}}_p^X=|{\tilde{t}}_p^X|e^{i\phi_p^X}$, where $|{\tilde{t}}_p^X|=(4[{\rm Re} \boldsymbol{\chi}^X(\Delta_p,\Delta_d)]^2+(1-2{\rm Im} \boldsymbol{\chi}^X(\Delta_p,\Delta_d))^2)^{1/2}$, and $\phi_p^X$ is in Eq.~\ref{presponse}. Both $|{\tilde{t}}_p^X|$ and $\phi_p^X$ are function of the probe and drive beam detunings. We can further define $\log {\tilde{t}}_p^X=\log |{\tilde{t}}_p^X|+i\phi_p^X$, where both $\log |{\tilde{t}}_p^X|$ and $\phi_p^X$ are real, and function of $\Delta_p,\Delta_d$. To investigate analyticity of $\log {\tilde{t}}_p^X$, we approximate $\boldsymbol{\chi}^X(\Delta_p,\Delta_d)$ for different 3LEs in the limit of $\Omega_p \to 0$ as
\bea
&&\boldsymbol{\chi}^{\Lambda}_{\rm a}=-\f{\Gamma_p}{\Delta_p+i\Gamma_{32}}\Big(1-\f{\Omega_d^2}{\Omega_d^2-(\Delta_p+i\Gamma_{21})(\Delta_p+i\Gamma_{32})}\Big),\nn \\
&&\boldsymbol{\chi}^{\Xi}_{\rm a}=\boldsymbol{\chi}^{\Xi}_l\Big(1-\f{\Omega_d^2}{\Omega_d^2-(\Delta_p+i\Gamma_{21}')(\Delta_p+i\Gamma_{31}')}\Big),\nn \\
&&\boldsymbol{\chi}^{V}_{\rm a}=\boldsymbol{\chi}^{V}_l\times\nn\\&&~~~\Big(1-\f{\Omega_d^2((\Delta_p+i\Gamma_{21}'+2i\Gamma_d)(\Delta_p+i\Gamma_{32}')-2\Omega_d^2)}{2(\Omega_d^2+\Gamma_d\Gamma_{31}')(\Omega_d^2-(\Delta_p+i\Gamma_{21}')(\Delta_p+i\Gamma_{32}'))}\Big),\nn
\eea
where $\boldsymbol{\chi}^{V}_l=\boldsymbol{\chi}^{\Xi}_l$. There are ordinary poles and branch point in the complex plane of $\log {\tilde{t}}_p^X$ coming respectively from vanishing denominator of $\boldsymbol{\chi}^X_a$ and ${\tilde{t}}_p^X=0$. We find in our numerics that all the poles of the complex function $\log {\tilde{t}}_p^X$ in the limit $\Omega_p \to 0$ lie in the lower half-plane of complex probe frequency detuning $\Delta_p$ for all three 3LEs. $\log {\tilde{t}}_p^X$ also decays faster than $1/|\Delta_p|$ as $|\Delta_p| \to \infty$ for all three 3LEs. We also observe the above two features of the complex function $\log {\tilde{t}}_p^X$ to hold in the presence of a non-zero non-radiative decay $\Gamma_{\gamma}$, which generates amplification of the probe beam in $\Lambda$ and $V$ systems. Thus, we get from the Kramers-Kronig relations:
\bea
&&\phi_p^X(\Delta_p,\Delta_d)=-\f{1}{\pi}\mathcal{P}\int_{-\infty}^{\infty}\f{\log |{\tilde{t}}_p^X(\Delta_p',\Delta_d)|}{\Delta_p'-\Delta_p}d\Delta_p',\label{KK1}\\
&&\log |{\tilde{t}}_p^X(\Delta_p,\Delta_d)|=\f{1}{\pi}\mathcal{P}\int_{-\infty}^{\infty}\f{\phi_p^X(\Delta_p',\Delta_d)}{\Delta_p'-\Delta_p}d\Delta_p'.\label{KK2}
\eea
Therefore, a knowledge of $\Delta_p$-dependence of $\phi_p^X$ ($|{\tilde{t}}_p^X|$) can be used to find $|{\tilde{t}}_p^X|$ ($\phi_p^X$) at any $\Delta_p$ employing Eq.~\ref{KK2} (Eq.~\ref{KK1}). We can perform these calculations in the presence and absence of the drive beam, and these amplitude and phase responses can be used to get the coherent amplification and the cross-Kerr phase shift in the limit of $\Omega_p \to 0$. In Fig.~\ref{kkfig}, we compare $\phi_p^X$ evaluated using the Kramers-Kronig relation in Eq.~\ref{KK1} to the exact $\phi_p^X$ for two different values $\Omega_p~(\la N_p\ra)$ in different 3LEs in the  absence and presence of a non-radiative decay. We find the values of $\phi_p^X$ obtained using the Kramers-Kronig relation match perfectly with the exact values for small $\Omega_p~(\la N_p\ra=0.01)$, but they differ when the value of $\Omega_p~(\la N_p\ra)$ increases. This mismatch at larger $\Omega_p$ is due to the breakdown of the Kramers-Kronig relations in the nonlinear regime of $\Omega_p$. Such breakdown is because of nonanalyticity of $\log {\tilde{t}}_p^X$ (appearance of poles in the upper half-plane) in the complex plane of probe frequency detuning at finite $\Omega_p$.  

\section{Discussion}
\label{dis}
We have analyzed scattering of two light beams from different configurations of a single 3LE embedded in an open waveguide. The treatment of both the beams within a fully quantum mechanical microscopic modeling at the few-photon regime and the inclusion of corresponding inherent relaxation are some unique features of our theory. Apart from the investigation of amplification and cross-Kerr interaction in this paper, such modeling can be useful for the study of electromagnetically induced transparency, nonreciprocity, quantum wave mixing, etc.

We here have compared our results with the recent experiments probing coherent amplification and cross-Kerr phase shift. For example, the values of the cross-Kerr phase shift in our study for a ladder system with both the probe and drive fields at the single-photon level are similar to the experimentally observed value of approximately 10 degrees in \textcite{HoiPRL2013}. We also find that the dependence of the cross-Kerr phase shift on the probe and drive power at a probe frequency that maximizes the phase shift shows similar trends as measured by \textcite{HoiPRL2013}. While \textcite{HoiPRL2013} applied a continuous probe and a drive pulse with a finite width in their experiment, our theoretical analysis is with continuous probe and drive beams. Nevertheless, the agreement in the value of the cross-Kerr phase shift in the experiment and our study is acceptable. Because the pulse width in the experiment is much longer than the relaxation time of the associated transition, and then our steady-state analysis with a continuum drive beam can be applied satisfactorily for a pulse drive beam of long width. Our prediction using the Heisenberg-Langevin equations approach for the maximum coherent amplification in a V also perfectly matches with that from a different analysis using the Lindblad equation in \textcite{AstafievPRL2010}.

It would now be interesting to experimentally test our other predictions on amplification and cross-Kerr nonlinearity for different types of 3LE, which have not yet been demonstrated in experiments. Notably, we have extended the theoretical analysis to the incoherent amplification, which was not previously measured in  the experiment by \textcite{AstafievPRL2010}, and we also here compared it to the coherent amplification. Finally, we have here correlated the amplitude and phase responses of the probe beam, which are separately used in experiments by \textcite{AstafievPRL2010} and \textcite{HoiPRL2013}. We mainly show that the knowledge of any one of the response can be employed to derive the other response for a weak probe beam. While the Kramers-Kronig relation seems to work correctly in the linear probe regime in Fig.~\ref{kkfig}, the application of the Kramers-Kronig relation in our study is in the nonlinear regime when both the probe and drive beams are considered. Therefore, our analysis opens up an exciting possibility for the examinations of amplification and cross-Kerr nonlinearity.

\appendix
\setcounter{figure}{0}
\renewcommand\thefigure{A\arabic{figure}}
\section{Linearization of photon dispersion}
\label{AppLin}
We begin with some arbitrary energy-momentum dispersions (e.g., $\om_{k\alpha},\om_{ck},\om_{dk},\om_{fk}$) of various photon and excitation modes, and write the full Hamiltonian for an X-type of 3LE in a 1D waveguide as
\bea
\f{\tilde{\mathcal{H}}_{X}^q}{\hbar}&=&\tilde{\om}_{21} \sigma^{\dg}\sigma+ \tilde{\om}_{31}\mu^{\dg}\mu+\int dk \Big(\sum_{\alpha=\pm}\om_{k\alpha}\varrho_{k\alpha}^{\dg}\varrho_{k\alpha}\nn\\&+&\om_{ck}c_{k}^{\dg}c_{k}+\om_{dk}d_{k}^{\dg}d_{k}+\om_{fk}f_{k}^{\dg}f_{k}+\lambda_2(c_k+c_k^{\dg})\sigma^{\dg}\sigma\nn\\&+&\lambda_3(f_k+f_k^{\dg})\mu^{\dg}\mu\Big)+ \f{\mathcal{H}^c_{X}}{\hbar},
\eea
where $\hbar\tilde{\om}_{21}$ and $\hbar\tilde{\om}_{31}$ are unscaled energies of level $|2\ra$ and $|3\ra$. Here, $\varrho_{k\alpha}^{\dg}$ are creation operators of photon modes of the probe ($\alpha=+$) and drive ($\alpha=-$) beams. All other operators and parameters are described in Sec.~\ref{model}. Nevertheless, it is convenient to assume a linear energy-momentum dispersion for the photons near the relevant transition frequency (e.g., $\om_{31}$ or $\om_{31}-\om_{21}$) of emitters for practical purposes as well as simplicity in theoretical treatments. Such linearization is done regularly in waveguide QED as discussed in \cite{RoyRMP2017}. In the process of linearization of probe and drive photon's energy, we divide the propagating photons as left-moving and right-moving photon modes with opposite momenta or wave-vectors (e.g., $\pm k_{0\alpha}$) which are related to some transition frequency $\om_{0\alpha}$. Thus, we write
\bea
&&\int\limits_{k_{\alpha}\simeq k_{0\alpha}}\om_{k\alpha}\varrho_{k\alpha}^{\dg}\varrho_{k\alpha} \simeq \int\limits_{k_{\alpha}\simeq k_{0\alpha}} [\om_{0\alpha}+v_{g\alpha}(k_{\alpha}-k_{0\alpha})]a_{k\alpha}^{\dg}a_{k\alpha},\nn\\
&&\int\limits_{k_{\alpha}\simeq -k_{0\alpha}}\om_{k\alpha}\varrho_{k\alpha}^{\dg}\varrho_{k\alpha} \simeq \int\limits_{k_{\alpha}\simeq -k_{0\alpha}} [\om_{0\alpha}-v_{g\alpha}(k_{\alpha}+k_{0\alpha})]b_{k\alpha}^{\dg}b_{k\alpha},\nn
\eea
where $a_{k\alpha}^{\dg}~[b_{k\alpha}^{\dg}]$ are creation operators for two different polarizations of right-moving [left-moving] photon modes of the probe and drive beams. Next, we extend the limit of integration of wave-vector $k_{\alpha}$ of left and right-moving photons from $-\infty$ to $\infty$. However, the contributions in these integrals come only from a narrow window around $\om_{0\alpha}$ (or $\pm k_{0\alpha}$) as most relevant physical processes in our studies occur for relatively small detuning of the incident or scattered photons from the relevant transition energy of the emitters. Similarly, we can linearize the dispersions of excitations of non-radiative decay and pure dephasing as follows: $\om_{dk}=(\om_{0-}-\om_{0+})+v_{gd}(k_d-k_{0d}), \om_{ck}=v_{gc}k$ and $\om_{fk}=v_{gf}k$, where we have only considered positive wave-vectors for stationary excitations. After the linearization of dispersions, we change the variables $k_{\alpha} \mp k_{0\alpha} \to k_{\alpha}$ and $k_d-k_{0d} \to k_d$. We also observe that the excitation numbers $\tilde{N}_1=N_1+\int_{-\infty}^{\infty}dk\: d_{k}^{\dg}d_{k}$ and $\tilde{N}_2=N_2-\int_{-\infty}^{\infty}dk\: d_{k}^{\dg}d_{k}$ are again conserved quantities for an $\Lambda$ system as they commute with $\tilde{\mathcal{H}}_{\Lambda}^q$. Thus, we redefine the linearized Hamiltonian with a proper set of $\tilde{N}_1$ and $\tilde{N}_2$ for an X-type of 3LE: 
\bea
\f{\mathcal{H}_{X}^q}{\hbar}&=&\f{\tilde{\mathcal{H}}_{X}^q}{\hbar}-\om_{0-}\tilde{N}_1-\om_{0+}\tilde{N}_2, \nn\\
&=&\om_{21}\sigma^{\dg}\sigma+ \om_{31}\mu^{\dg}\mu+\int_{-\infty}^{\infty}dk \: \big[\sum_{\alpha=\pm}v_{g\alpha}k(a_{k\alpha}^{\dg}a_{k\alpha}\nn\\&-&b_{k\alpha}^{\dg}b_{k\alpha})+v_{gd}kd_{k}^{\dg}d_{k}+v_{gc}kc_{k}^{\dg}c_{k}+v_{gf}kf_{k}^{\dg}f_{k}\nn\\&+&\lambda_2(c_k+c_k^\dg)\sigma^\dg\sigma+\lambda_3(f_k+f_k^\dg)\mu^\dg\mu \big]+\f{\mathcal{H}^c_X}{\hbar},\label{scHam}
\eea
where we have for example, $\om_{21}=\tilde{\om}_{21}-(\om_{0-}-\om_{0+}),\om_{31}=\tilde{\om}_{31}-\om_{0-}$ for an $\Lambda$ system. Substituting $v_{g+}=v_{g-}=v_{gd}=v_{gc}=v_{gf}=v_g$ in Eq.~\ref{scHam} for simplicity, we get the Hamiltonian in Eq.~\ref{Ham1Q}.

\section{Heisenberg-Langevin equations for an $\Lambda$ system}
\label{AppHL}
We apply the Heisenberg-Langevin equations approach to calculate the time-evolution of the emitter and light fields after they interact. We get time derivative of all operators associated with the emitter and light and excitation fields using the Heisenberg equation. For example, the Heisenberg equation for the right-moving probe field $a_{k+}$ in an $\Lambda$-type 3LE in waveguide using Eq.~\ref{Ham1Q} is
\bea
\f{da_{k+}(t)}{dt}=-i v_gk a_{k+}(t)-i g_p\mu(t),
\eea
which can be integrated over time with an initial value $a_{k+}(t_0)$ to obtain 
\bea
a_{k+}(t)=a_{k+}(t_0)e^{-iv_gk(t-t_0)}-ig_p\int_{t_0}^{t}dt'e^{-iv_gk(t-t')}\mu(t').\nn\\\label{b2}
\eea 
Similarly, we derive for other operators of the propagating light fields and excitations: 
\bea
a_{k-}(t)&=&a_{k-}(t_0)e^{-iv_gk(t-t_0)}-ig_d\int_{t_0}^{t}dt'e^{-iv_gk(t-t')}\nu^\dg(t'),\nn\\
b_{k+}(t)&=&b_{k+}(t_0)e^{-iv_gk(t-t_0)}-ig_p\int_{t_0}^{t}dt'e^{-iv_gk(t-t')}\mu(t'),\nn\\
b_{k-}(t)&=&b_{k-}(t_0)e^{-iv_gk(t-t_0)}-ig_d\int_{t_0}^{t}dt'e^{-iv_gk(t-t')}\nu^\dg(t'),\nn\\
d_{k}(t)&=&d_{k}(t_0)e^{-iv_gk(t-t_0)}-i\gamma \int_{t_0}^{t}dt'e^{-iv_gk(t-t')}\sigma(t'),\nn\\
c_{k}(t)&=&c_{k}(t_0)e^{-iv_gk(t-t_0)}-i\lambda_2 \int_{t_0}^{t}dt'e^{-iv_gk(t-t')}\sigma^\dg(t')\sigma(t'),\nn\\
f_{k}(t)&=&f_{k}(t_0)e^{-iv_gk(t-t_0)}-i\lambda_3 \int_{t_0}^{t}dt'e^{-iv_gk(t-t')}\mu^\dg(t')\mu(t').\nn\\\label{b3}
\eea
Next, we write the Heisenberg equation for the operators of the 3LE. For example,
\bea
\f{d\sigma\sigma^\dg}{dt}&=&\int_{-\infty}^{\infty}dk[-ig_d((a^\dg_{k-}(t)+b^\dg_{k-}(t))\nu^\dg(t)-\nu(t)(a_{k-}(t)\nn\\&+&b_{k-}(t)))+i\gamma(\sigma^\dg(t) d_k(t)-d_k^\dg(t)\sigma(t))].\label{si}
\eea
We employ the formal solutions of the photon and excitation field operators from Eqs.~\ref{b2},\ref{b3} to integrate the right side of Eq.~\ref{si}. Thus, we get
\bea
&&ig_d\int_{-\infty}^{\infty}dk\:a_{k-}(t)=\int_{-\infty}^{\infty}dk[ig_da_{k-}(t_0)e^{-iv_gk(t-t_0)}\nn\\&&~~~~~~~~~~~~~~~~~~~~~~~~~~~~+g_d^2\int_{t_0}^{t}dt'e^{-iv_gk(t-t')}\nu^\dg(t')]\nn\\
&&~~~~~~~~~~~~=i\eta_{a-}(t)+g_d^2\int_{t_0}^{t}dt'2\pi \delta(v_g(t-t'))\nu^\dg(t')\nn\\
&&~~~~~~~~~~~~=i\eta_{a-}(t)+\Gamma_d\nu^\dg(t),\label{no1}
\eea
where $\Gamma_d=\pi g_d^2/v_g$ is a measure of relaxation of the emitter due to the right-moving drive field, and $\eta_{a-}(t)=\int_{-\infty}^{\infty}dk\:g_de^{-iv_gk(t-t_0)}a_{k-}(t_0)$ is a noise appears due to the integration of the corresponding drive field. We can carry out the other integration in Eq.~\ref{si} to find
\bea
ig_d\int_{-\infty}^{\infty}dk\:b_{k-}(t))&=&i\eta_{b-}(t)+\Gamma_d\nu^\dg(t),\label{no2}\\
i\gamma \int_{-\infty}^{\infty}dk\:d_{k}(t)&=&i\eta_{d}(t)+\Gamma_{\gamma}\sigma(t),\label{no3}
\eea
where the relaxation rate $\Gamma_\gamma=\pi \gamma^2/v_g$, and  the noise terms $\eta_{b-}(t)=g_d\int_{-\infty}^{\infty}dk\:e^{-iv_gk(t-t_0)}b_{k-}(t_0)$, and $\eta_{d}(t)=\gamma\int_{-\infty}^{\infty}dk\:e^{-iv_gk(t-t_0)}d_{k}(t_0)e^{-iv_gk(t-t_0)}$. 
Using Eqs.~\ref{no1},\ref{no2},\ref{no3} in Eq.~\ref{si}, we get the following quantum Langevin equation:
\bea
\f{d\sigma\sigma^\dg}{dt}=&&i(\nu(\eta_{a-}+\eta_{b-})-(\eta_{a-}^\dg+\eta_{b-}^\dg)\nu^\dg)+i(\sigma^\dg\eta_{d}\nn\\&&-\eta^\dg_{d}\sigma)+4\Gamma_d\nu\nu^\dg+2\Gamma_\gamma \sigma^\dg\sigma. \label{no4}
\eea
Next, we take expectation of the above Eq.~\ref{no4} in the initial state $|\psi\ra$ and employ the action of initial photon and excitation fields on $|\psi\ra$ given in Sec.~\ref{3}. Thus, we get
\bea
\f{d\la\sigma\sigma^\dg\ra}{dt}=&&i\Omega_{d}(\la\nu\ra e^{-i\om_d(t-t_0)}-\la\nu^\dg\ra e^{i\om_d(t-t_0)})\nn\\&&+4\Gamma_d\la\nu\nu^\dg\ra+2\Gamma_\gamma(1- \la\sigma\sigma^\dg\ra-\la\nu\nu^\dg\ra),
\eea
which is one of the eight equations in Eq.~\ref{eom3LEq} when written in terms of the variables of $\mathcal{M}_{\Lambda}$. We find from the above equation that the net decay rate to the ground state $|1\ra$ is $4\Gamma_d$ from level $|3\ra$ due to the transition coupling by the drive field and $2\Gamma_\gamma$ from level $|2\ra$ due to non-radiative decay. We can obtain similar equations for all other variables of $\mathcal{M}_{\Lambda}$ to get the matrix Eq.~\ref{eom3LEq}.

\section{Derivation of transmission and reflection coefficients for an $\Lambda$ system}
\label{AppDt}
We derive the total transmitted and reflected power for both the probe and drive beams by summing the equation for power spectrum in Eq.~\ref{power1},\ref{power2} over all frequencies. We find from Eq.~\ref{power1} for the transmitted power:
\bea
&&\int_{-\infty}^{\infty}d\om P_{\rm tr,\alpha}(t,\om)=\la a_{x\alpha}^{\dg}(t)a_{x\alpha}(t)\ra, \label{pow3}\\
&&a_{x\alpha}(t)=\int_{-\infty}^{\infty} \f{dk}{\sqrt{2\pi}}\:e^{ikx}a_{k\alpha}(t).\label{f1}
\eea
Replacing $a_{k+}(t)$ from Eq.~\ref{b2} in Eq.~\ref{f1}, we get for $x>0$
\bea
a_{x+}(t)=\f{\eta_{a+}(t-\f{x}{v_g})}{\sqrt{2\pi}}-i\f{g_p\sqrt{2\pi}}{v_g}\mu(t-\f{x}{v_g}).\label{f2}
\eea
Using Eq.~\ref{f2} in Eq.~\ref{pow3}, we get total transmitted probe power at some position $x>0$ 
\bea
\la a_{x+}^{\dg}(t)a_{x+}(t)\ra_{x>0}=&&\f{\Om_p^2}{2\Gamma_{p}v_g}-\f{2\Om_p}{v_g}{\rm Im}[\la\mu^\dg(t')\ra e^{-i\om_pt'}]\nn\\
&&+\f{2\Gamma_p}{v_g}\la\mu^\dg(t')\mu(t')\ra,
\eea
where $t'=t-x/v_g$. Dividing the total transmitted probe power by the incident probe intensity, we get the transmission coefficient of the probe beam at $x=0+$:
\bea
\mathcal{T}^{\Lambda}_p(t)&=&1+\f{4\Gamma_p^2}{\Omega_p^2} \mathcal{N}_{2}(t)+\f{4\Gamma_p}{\Omega_p} {\rm Im}[\mathcal{M}_{1}(t)].
\eea
As before, we can write for the left-moving mode of the probe beam at $x<0$:
\bea
b_{x+}(t)=\f{\eta_{b+}(t-\f{x}{v_g})}{\sqrt{2\pi}}-i\f{g_p\sqrt{2\pi}}{v_g}\mu(t-\f{x}{v_g}).
\eea
Thus, the total reflected probe power at some $x<0$ is given by
\bea
\la b_{x+}^{\dg}(t)b_{x+}(t)\ra_{x<0}=\f{2\Gamma_p}{v_g}\la\mu^\dg(t')\mu(t')\ra.
\eea
Thus, we find for the reflection coefficient of the probe beam at $x=0-$ after dividing the above expression by the incident probe intensity:
\bea
\mathcal{R}^{\Lambda}_p(t)&=&\f{4\Gamma_p^2}{\Omega_p^2} \mathcal{N}_{2}(t).
\eea

\section{Transport properties of V-type 3LE}
\label{App1}
\bea
&&\f{d\boldsymbol{\mathcal{M}}_{V}}{dt}=\boldsymbol{\mathcal{R}}_{V}\boldsymbol{\mathcal{M}}_{V}+\boldsymbol{\Omega}_{V},~~{\rm where}~~\boldsymbol{\mathcal{R}}_{V}=\label{eom4LEq}\\
&&\left(
\scalemath{.9}{\begin{array}{cccccccc}
	\kappa _4 & 0 & 0 & i \Omega _d & -i \Omega _d & i \Omega _p & 0 & 0 \\
	0 & \kappa _5 & i \Omega _d & -i \Omega _p & -2 i \Omega _p & 0 & 0 & 0 \\
	0 & i \Omega _d & \kappa _6 & 0 & 0 & 0 & 0 & -i \Omega _p \\
	i \Omega _d & 0 & 0 & -4 \Gamma _d-2 \Gamma _{\gamma } & 0 & 0 & 0 & -i \Omega _d \\
	-i \Omega _d & -i \Omega _p & 0 & 4 \Gamma _d-4 \Gamma _p & -4 \Gamma _p & 0 & i \Omega _p & i \Omega _d \\
	i \Omega _p & 0 & 0 & 0 & 0 & \kappa _6^* & -i \Omega _d & 0 \\
	0 & 0 & 0 & i \Omega _p & 2 i \Omega _p & -i \Omega _d & \kappa _5^* & 0 \\
	0 & 0 & -i \Omega _p & -i \Omega _d & i \Omega _d & 0 & 0 & \kappa _4^* \\
\end{array}}
\right),\nn
\eea
$\boldsymbol{\mathcal{M}}_{V}(t)=(\mathcal{N}^*_3,\mathcal{S}_3,\mathcal{M}^*_3,\mathcal{N}_4,\mathcal{S}_4,\mathcal{M}_3,\mathcal{S}^*_3,\mathcal{N}_3)^{T}$ and $\boldsymbol{\Omega}_{V}=(0,i\Omega_p,0,0,4\Gamma_{p},0,-i\Omega_p, 0)^{T}$. We define the diagonal entries $\kappa_4=-i\Delta_d-\Gamma_\gamma-2\Gamma_d,~\kappa_5=-i\Delta_p-2\Gamma_{p},~\kappa_6=\kappa_4^*+\kappa_5,~\Delta_p=\om_{21}-\om_p,~\Delta_d=\om_{31}-\om_d$. We have used the following definitions for the expectation of the emitter's operators in $\boldsymbol{\mathcal{M}}_{V}(t)$:
\bea
\mathcal{N}_3(t)&=&\la \psi |\nu(t)| \psi \ra e^{-i\om_d(t-t_0)}, \nn\\
\mathcal{S}_3(t)&=&\la \psi |\sigma(t)| \psi \ra e^{i\om_p(t-t_0)},\nn\\
\mathcal{S}_4(t)&=&\la \psi|\sigma(t)\sigma^{\dg}(t)| \psi \ra,\nn\\
\mathcal{M}_3(t)&=&\la \psi |\mu(t)| \psi \ra e^{i(\om_d-\om_p)(t-t_0)}, \nn\\
\mathcal{N}_4(t)&=&\la \psi |\nu(t)\nu^{\dg}(t)| \psi \ra.\nn
\eea
\bea
\mathcal{T}^V_p(t)&=&1+\f{4\Gamma_p^2}{\Omega_p^2} (1-\mathcal{N}_{4}(t)-\mathcal{S}_{4}(t))\nn\\&-&\f{4\Gamma_p}{\Omega_p} {\rm Im}[\mathcal{S}_{3}^*(t)],\label{transp2}\\
\mathcal{T}^V_d(t)&=&1+\f{4\Gamma_d^2}{\Omega_d^2} \mathcal{N}_{4}(t)-\f{4\Gamma_d}{\Omega_d} {\rm Im}[\mathcal{N}_{3}(t)],\label{transd2}\\
\mathcal{R}^V_p(t)&=&\f{4\Gamma_p^2}{\Omega_p^2} (1-\mathcal{N}_{4}(t)-\mathcal{S}_{4}(t)),\\~\mathcal{R}^V_d(t)&=&\f{4\Gamma_d^2}{\Omega_d^2} \mathcal{N}_{4}(t).\label{refd2}
\eea
\section{Transport properties of ladder($\Xi$)-type 3LE}
\label{App2}
\bea
&&\f{d\boldsymbol{\mathcal{M}}_{\Xi}}{dt}=\boldsymbol{\mathcal{R}}_{\Xi}\boldsymbol{\mathcal{M}}_{\Xi}+\boldsymbol{\Omega}_{\Xi},~~{\rm where}~~\boldsymbol{\mathcal{R}}_{\Xi}=\label{eom5LEq}\\
&&\left(
\scalemath{.9}{
	\begin{array}{cccccccc}
		\kappa _ 8 & -i \Omega _d & 0 & 0 & 0 & i \Omega _p & 0 & 0 \\
		-i \Omega _d & \kappa _ 5 & 0 & -i \Omega _p & -2 i \Omega _p & 0 & 0 & 0 \\
		0 & 0 & \kappa _ 7^* & -2 i \Omega _d & -i \Omega _d & 0 & 0 & -i \Omega _p \\
		0 & 0 & -i \Omega _d & -4 \Gamma _d-2 \Gamma _{\gamma } & 0 & i \Omega _d & 0 & 0 \\
		0 & -i \Omega _p & 0 & 2 \Gamma _{\gamma }-4 \Gamma _p & -4 \Gamma _p & 0 & i \Omega _p & 0 \\
		i \Omega _p & 0 & 0 & 2 i \Omega _d & i \Omega _d & \kappa _ 7 & 0 & 0 \\
		0 & 0 & 0 & i \Omega _p & 2 i \Omega _p & 0 & \kappa _ 5^* & i \Omega _d \\
		0 & 0 & -i \Omega _p & 0 & 0 & 0 & i \Omega _d & \kappa _ 8^* \\
	\end{array}}
\right),\nn
\eea
$\boldsymbol{\mathcal{M}}_{\Xi}(t)=(\mathcal{N}^*_5,\mathcal{S}_5,\mathcal{M}^*_5,\mathcal{N}_6,\mathcal{S}_6,\mathcal{M}_5,\mathcal{S}^*_5,\mathcal{N}_5)^{T}$ and $\boldsymbol{\Omega}_{\Xi}=(0,i\Omega_p,i\Omega_d,0,4\Gamma_{p},-i\Omega_d,-i\Omega_p, 0)^{T}$. We define the diagonal entries $\kappa_7=-i\Delta_d-\tilde{\Gamma},~\kappa_5=-i\Delta_p-2\Gamma_{p},~\kappa_8=\kappa_7-\kappa_5^*,~\Delta_p=\om_{21}-\om_p,~\Delta_d=\om_{31}-\om_{21}-\om_d$. We have used the following definitions for the expectation of the emitter's operators in $\boldsymbol{\mathcal{M}}_{\Xi}(t)$:
\bea
\mathcal{N}_5(t)&=&\la \psi |\nu(t)| \psi \ra e^{-i(\om_d+\om_p)(t-t_0)}, \nn\\
\mathcal{S}_5(t)&=&\la \psi |\sigma(t)| \psi \ra e^{i\om_p(t-t_0)},\nn\\
\mathcal{S}_6(t)&=&\la \psi|\sigma(t)\sigma^{\dg}(t)| \psi \ra,\nn\\
\mathcal{M}_5(t)&=&\la \psi |\mu(t)| \psi \ra e^{i\om_d(t-t_0)}, \nn\\
\mathcal{N}_6(t)&=&\la \psi |\nu(t)\nu^{\dg}(t)| \psi \ra.\nn
\eea
\bea
\mathcal{T}^{\Xi}_p(t)&=&1+\f{4\Gamma_p^2}{\Omega_p^2} (1-\mathcal{N}_{6}(t)-\mathcal{S}_{6}(t))\nn\\&-&\f{4\Gamma_p}{\Omega_p} {\rm Im}[\mathcal{S}_{5}^*(t)],\label{transp3}\\
\mathcal{T}^{\Xi}_d(t)&=&1+\f{4\Gamma_d^2}{\Omega_d^2} \mathcal{N}_{6}(t)-\f{4\Gamma_d}{\Omega_d} {\rm Im}[\mathcal{M}_{5}^*(t)],\label{transd3}\\
\mathcal{R}^{\Xi}_p(t)&=&\f{4\Gamma_p^2}{\Omega_p^2} (1-\mathcal{N}_{6}(t)-\mathcal{S}_{6}(t)),\\\mathcal{R}^{\Xi}_d(t)&=&\f{4\Gamma_d^2}{\Omega_d^2} \mathcal{N}_{6}(t).\label{refd3}
\eea

\section{Quantum vs. classical drive beam}
\label{App3}
In most earlier studies of such 3LEs with two beams in an open waveguide, the drive light is considered to be a classical beam at a relatively higher intensity \cite{RoyRMP2017}. The Hamiltonian in Eqs.~\ref{Ham1Q} and \ref{HamLam} for a classical drive beam can be rewritten in a frame rotating at the drive frequency $\om_d$ as
\bea
\f{\mathcal{H}_{\Lambda}^{cl}}{\hbar}&=&\om_{21} \sigma^{\dg}\sigma+ (\om_{31}+\om_d)\mu^{\dg}\mu+\Omega_d(\nu+\nu^{\dg})\nn\\&+&\int_{-\infty}^{\infty}dk \big[v_gk(a_{k+}^{\dg}a_{k+}-b_{k+}^{\dg}b_{k+}+c^{\dg}_kc_k+d^{\dg}_kd_k\nn \\&+&f^{\dg}_kf_k)+(g_p\mu^{\dg}\beta_{k+}+\gamma \sigma^{\dg} d_k+h.c.)\nn\\&+&\lambda_2(c_k+c_k^{\dg})\sigma^{\dg}\sigma +\lambda_3(f_k+f_k^{\dg})\mu^{\dg}\mu\big],\label{Ham1C}
\eea
where  $\Omega_d$ is the Rabi frequency of the drive beam. Such classical modeling of the drive beam ignores any relaxation induced by the beam to the optical transition.  The explicit inclusion of relaxation in microscopic quantum modeling causes differences in the probe beam lineshapes from a driven $\Lambda$-type 3LE obtained by classical and quantum modeling at a weak intensity of the drive beam. Such differences are relatively less significant for a $V$ and a ladder system.

By setting the decay $\Gamma_d=0$ in $\boldsymbol{\mathcal{R}}_{\Lambda}$ of Eq.~\ref{eom3LEq}, we get the time-evolution of a $\Lambda$-type 3LE and a probe beam for a classical drive beam of strength $\Omega_d$ as in Eq.~\ref{Ham1C}. The relaxation terms with $\Gamma_d$ in Eq.~\ref{eom3LEq} appear due to microscopic quantum modeling of the drive beam in Eqs.~\ref{Ham1Q} and \ref{HamLam},  and it has introduced an off-diagonal relaxation term in the evolution matrix $\boldsymbol{\mathcal{R}}_{\Lambda}$ apart from the extra relaxation in the diagonal entries of $\boldsymbol{\mathcal{R}}_{\Lambda}$. This off-diagonal relaxation term in $\boldsymbol{\mathcal{R}}_{\Lambda}$ generates some interesting differences in the lineshapes and power spectra of the scattered probe light for a weak drive in Eq.~\ref{Ham1Q} in comparison to a weak drive beam in Eq.~\ref{Ham1C}. While a weak probe beam (single-photon regime) is perfectly reflected when $\Gamma_{\gamma}=0$ and $\Omega_d \to 0$ for classical modeling of the drive beam, it can be fully transmitted as $\Omega_d \to 0$ for quantum modeling. In the absence of the classical drive beam ($\Omega_d \to 0$) in Eq.~\ref{Ham1C} and $\Gamma_{\gamma}=0$, the $\Lambda$-type 3LE reduces to an effective 2LE with a transition $|2\ra \leftrightarrow |3\ra$, which is coupled to the probe beam. Therefore, a probe photon at resonant $(\Delta_p=0)$ to this transition perfectly reflects in a side-coupled waveguide QED system. On the other hand, for quantum modeling of drive beam in Eqs.~\ref{Ham1Q} and \ref{HamLam}, there would be some spontaneous emission from the excited $|3\ra$ to $|1\ra$ even when $\Omega_d \to 0$. This is due to the off-diagonal relaxation term arising from the coupling of $|3\ra  \leftrightarrow |1\ra$ transition to the vacuum modes of the drive beam. Such spontaneous emission brings the population of the 3LE to $|1\ra$ level by emitting a drive photon of $-$ polarization.  Nevertheless, the conversion of probe photon to drive photon of different polarization can occur for maximum a single photon at $\Omega_d \to 0$ in Eqs.~\ref{Ham1Q} and \ref{HamLam}, and it is a transient process as the probe field does not further interact with the emitter. Therefore, probe photons fully transmit through the $\Lambda$-type 3LE at long-time steady-state when $\Omega_d \to 0$.

\section*{Acknowledgments}
We thank C. M. Wilson for discussion. D.R. gratefully acknowledges the funding from the Department of Science and Technology, India via the Ramanujan Fellowship.

\bibliography{bibliography2}
\end{document}